\documentclass[aps,preprintnumbers,floats, prd, twocolumn, longbibliography, nofootinbib]{revtex4-1}
\usepackage{graphicx}
\usepackage{dcolumn}
\usepackage{bm}
\usepackage{amsmath}
\usepackage{amsthm}
\usepackage{multirow}
\usepackage{enumerate}
\usepackage{amsfonts}
\usepackage{ifthen}
\usepackage{psfrag}
\usepackage{slashed}
\usepackage{hyperref}
\usepackage{gensymb}
\usepackage[utf8]{inputenc}
\usepackage{color}
\usepackage{subfigure}
\usepackage{ulem}
\usepackage[utf8]{inputenc}
\usepackage{float}
\usepackage{verbatim}

\newcommand{\be}{\begin{equation}}
\newcommand{\ee}{\end{equation}}
\newcommand{\bea}{\begin{eqnarray}}
\newcommand{\eea}{\end{eqnarray}}
\newcommand{\met}{\slashed{E}_T}

\newcommand{\iab}{\rm ab^{-1}}

\begin{document} 

\title{Heavy Neutrino Searches via Same-sign Lepton Pairs at the Higgs Factory}

\author{Yu Gao$^{1}$}
\email{gaoyu@ihep.ac.cn}
\author{Kechen Wang$^{2}$}
\email{kechen.wang@whut.edu.cn}

\affiliation{$^1$Key Laboratory of Particle Astrophysics, Institute of High Energy Physics, Chinese Academy of Sciences, Beijing, 100049, China}
\affiliation{$^2$Department of Physics, School of Science, Wuhan University of Technology, 430070 Wuhan, Hubei, China}

\begin{abstract}
This paper investigates the $e^-e^+\rightarrow Zh_1$ sensitivity for Higgs boson's rare decay into heavy neutrinos $h_1 \rightarrow NN$ at the proposed electron-positron collider, with the focus on multi-lepton final states that contain same-sign lepton pairs. $h_1 \rightarrow NN$ decay can derive from Higgs boson's mixing with new physics scalar(s) that is complementary to the contribution from active-sterile neutrino mixings. 
We consider the scenario with a singlet scalar which gives the heavy neutrino mass and has a small mixing with the SM Higgs boson.
We analyze the semileptonic, fully leptonic and mixed $NN$ decay scenarios, and categorize the signal on the number of leptons in the final state: $ \ell^\pm \ell^\pm$ with at least 3 jets, $\ell^\pm \ell^\pm \ell$ with at least 2 jets, and $e^\pm e^\pm \mu^\mp \mu^\mp$ plus with at least 1 jet, each containing one or two same-sign dilepton system(s). Selection cuts are optimized for the presence of the associated $Z$-boson, which leads to additional backgrounds at the $e^-e^+$ collider. The Standard Model background channels are systematically analyzed. Sensitivity limits on $h_1 \rightarrow NN$ branching fractions are derived for signals with 2-4 final leptons assuming the heavy neutrino masses are between 10 and 60 GeV. With 240 GeV center-of-mass energy and 5.6 ab$^{-1}$ design luminosity, $h_1 \rightarrow NN$ branching fraction can be probed to $2\times 10^{-4}$ in $2\ell$ and $3\ell$ channels, and $6\times 10^{-4}$ in the $4\ell$ channel at $95\%$ confidence level. $3\ell, 4\ell$ channels expect one or fewer background event, and their sensitivities saturate the statistic limit at 5.6 ab$^{-1}$ luminosity. A same-sign trilepton ($\ell^\pm \ell^\pm \ell^\pm$) signal in the $3\ell$ channel is also discussed. Our search strategy provides an approach to discovering the singlet scalar and exploring the origin of neutrino masses at future $e^-e^+$ colliders.
\end{abstract}

\maketitle

\section{Introduction}
\label{sect:intro}

The collider search for massive neutrinos plays an important role in the testing of neutrino mass models that base on the seesaw mechanism~\cite{Minkowski:1977sc,Yanagida:1979as,Mohapatra:1979ia,Glashow:1979nm,GellMann:1980vs}. Mass of the active neutrino $\nu$ is generated by mixing the left-handed neutrino ($\nu_L$) of the Standard Model (SM) with the additional right-handed neutrino $N_R$, resulting in a heavy mass eigenstate $N$ that has a small SM $\nu_L$ component. The heavy $N$ acquires effective couplings to the SM model gauge bosons via its weakly charged $\nu_L$ component~\cite{Atre:2009rg} and is extensively searched at colliders. See the review~\cite{Deppisch:2015qwa,Cai:2017mow} and references therein for collider searches of heavy neutrinos.  Experimental limits can be found in Refs.~\cite{Sirunyan:2018mtv, CMS:2018jxx, SHiP:2018xqw, ATLAS:2019kpx, LHCb:2020wxx, CMS:2021lzm}. 

Recent phenomenology studies on heavy neutrino searches at $e^-e^+$ colliders vary according to its different production and decay channels.
For the prompt decays of heavy neutrinos, searches has been investigated under the symmetry protected seesaw model in Refs.~\cite{Antusch:2015rma, Antusch:2015mia, Antusch:2016ejd} and the EFT framework in Ref.~\cite{Barducci:2020icf};
heavy neutrinos from the production process $e^- e^+ \to \nu N$ are studied at center-of-mass energy $\sqrt{s} = 240$ GeV~\cite{Liao:2017jiz} and $Z-$pole running mode~\cite{Ding:2019tqq, Blondel:2021mss};
Ref.~\cite{Asaka:2016rwd} considers the $N$ production from $B-$meson decays $B^+ \to \mu^+ N \to \mu^+ \mu^+ \pi^-$, while
Refs.~\cite{Banerjee:2015gca, Zhang:2018rtr} consider the lepton number violating final state for the Majorana $N$ from the process $e^-e^+ \to \ell^\pm W^\mp N \to \ell^\pm \ell^\pm + 4 j$.
For the displaced vertex searches of long-lived $N$, Refs.~\cite{Antusch:2016vyf,Wang:2019xvx} consider the production from $e^- e^+ \to \nu N$, while Ref.~\cite{Dib:2019tuj} considers the production from tau decays at $B$-factories.
Furthermore, limits have been also set indirectly based on the $N$'s corrections to the decay branching ratio of $Z-$boson into two leptons Br$(Z \to \ell_1^\mp \ell_2^\pm )$ at $Z-$pole running mode~\cite{Abada:2014cca} and to the cross section of the SM Higgs boson production process $e^-e^+\rightarrow W^-W^+h_1$ at $\sqrt{s} = 3$ TeV~\cite{Baglio:2017fxf}.

While the heavy $N_R$ mass scale explains the tiny active neutrino mass, it also suppresses the left-right neutrino mixing, and makes weak production of heavy neutrinos difficult when this mixing is small. Alternatively, heavy neutrinos can also be produced in case they couple to Beyond the Standard Model (BSM) mediators, e.g. extra gauge bosons or scalars that couple to SM particles. Currently, such BSM gauge bosons are stringently constrained by resonance searches~\cite{Sirunyan:2018exx, Aad:2019hjw} and electroweak precision data~\cite{Langacker:2008yv, Akhmedov:2013hec}. In comparison, a mixing between the Higgs boson with BSM singlet scalar(s) is less constrained~\cite{DiLuzio:2017tfn}, and it is among the major physics goals at future Higgs factories~\cite{DiMicco:2019ngk, CEPCStudyGroup:2018ghi}.

The right-handed neutrino can obtain its mass by coupling to BSM scalars with a non-zero vacuum expectation value (vev). Generally such scalars mix with the SM Higgs doublet scalar, so if kinematically allowed, the physical Higgs boson can decay into heavy neutrinos through its BSM component. Since this decay occurs directly through the scalar mixing, it is insensitive to $\nu_L-N_R$ mixing, thus it is complementary to $|V_{\ell N}|^2$ based searches. Typical implementations involve extending the SM's scalar sector, e.g. from UV-complete models such as left-right symmetric models~\cite{Wyler:1982dd}, $U(1)_{B-L}$ models~\cite{Mohapatra:1979ia}, next to minimal supersymmetric model~\cite{Fayet:1974pd}, or alternatively, rising from effective theory operators~\cite{Graesser:2007yj}, 4th generation neutrino~\cite{Belotsky:2002ym}, etc.

When the Higgs boson decay into heavy neutrinos $h_1 \rightarrow N N$, a rare multi-lepton Higgs decay emerges. $N$ subsequently decays to SM particles through its $\nu_L$ component's weak interaction. The fully leptonic $N\rightarrow \ell\ell'\bar{\nu}$ and semileptonic $N\rightarrow \ell jj$ channels are interesting at collider searches due to the presence of measurable charged leptons in the final state. When $N$ is a Majorana fermion, semileptonic $NN$ decay leads to the characteristic lepton-number violating (LNV) same-sign (SS) dilepton, as recently studied as a LNV probe for Higgs-BSM scalar mixing~\cite{Shoemaker:2010fg, Maiezza:2015lza, Nemevsek:2016enw,Moretti:2019yln}. 

For collider searches, increased lepton multiplicity and the existence of SS
lepton pairs greatly reduce the SM's background, particularly for hadron collisions.
In our previous work~\cite{Gao:2019tio}, we demonstrate that a characteristic signal of two same-sign same-flavor (SSSF) lepton pairs plus missing energy from the $NN$ decay can probe the Higgs-singlet mixing with high precisions at the current and future $pp$ colliders. 
At lepton colliders, in comparison, hadronic backgrounds are controllable and the dominant Higgs boson production is the $e^-e^+\rightarrow Zh_1$ channel.

The associated $Z$-boson always appears and it provides additional leptons or jets to the final state. 
This is a complication at the lepton collider. Lepton colliders are known for high sensitivity to relatively soft leptons, yet intrinsic multi-tau background for SS dileptons do exist. Thus, it is of interest to systemically study the multi-lepton sensitivity at future lepton colliders, and we will show that the associated $Z$-boson leads to both signal and extra background. In addition, a characteristic SS trilepton signal emerges, unique to the dominant $Zh_1$ channel at the $e^-e^+$ Higgs factory. 
In this consecutive work, according to the semileptonic, fully leptonic and mixed $NN$ decays we categorize the signal on the number of final state leptons at future $e^- e^+$ colliders. After simulating the signal and SM background processes and performing the data analyses, we forecast the sensitivity limits on $h_1 \to NN$ branching fractions for signals with $2-4$ final leptons assuming the heavy neutrino masses are between 10 and 60 GeV. Our signal channels at both $pp$ and $e^-e^+$ colliders provide approaches to discovering the singlet scalar and exploring the origin of neutrino masses.  

The proposed lepton collider missions, e.g. the CEPC~\cite{CEPCStudyGroup:2018rmc}, ILC~\cite{Baer:2013cma} and FCC-ee~\cite{Abada:2019zxq}, are designed to yield ${\cal O} (10^{6-7})$ Higgs events. 
Any Higgs decay branching limit would be statistically capped by the collider's luminosity-inverse. 
This study on the relevant backgrounds would also
help understanding whether future $h_1 \rightarrow NN$ sensitivity would saturate luminosity limits.

This paper is organized as follows: Section~\ref{sect:singlet_extension} briefly discusses a minimal singlet extension to the SM that implements the $h_1 \rightarrow NN$ decay channel. Section \ref{sect:semileptonic} to \ref{sect:fullyleptonic} categorize signal channels on the number of final state leptons, analyze each channel's SM background and the event selection strategies. In Section~\ref{sect:summary} we give the sensitivity limits at future Higgs factory.

\section{Model Setup}
\label{sect:singlet_extension}

For collider search purposes, we adopt the phenomenological simple extension to the Standard Model that implements a Type-I seesaw mechanism.  With a scalar $S$ and Majorana fermion $N_R$ that are both SM gauge singlets, the addition to interaction Lagrangian is given by,
\bea 
&\cal L & \supset\ y_D \bar{L} \tilde{H} N_R+ y_S S \bar{N}_R^C N_R + h.c. \nonumber \\
 & & ~+ \frac{\lambda}{2}\, |H|^2|S|^2 +V_{S}.
\label{eqn:Lag}
\eea
where $\tilde{H}=i\sigma_2 H^*$ and $H$ is the SM Higgs doublet, $y_D$ and $y_S$ are the couplings that give the Dirac and Majorana mass terms after the doublet $H$ and the singlet $S$ obtain their vacuum expectation values 
$v_H = $ 246 GeV
and $v_S$.
The light neutrino mass eigenstate(s) obtain a mass $m_\nu\sim y_D^2v_H^2/(y_S v_S)$ and the left-right neutrino mixing
parameter $|V_{\ell N}| \sim \sqrt{m_\nu/m_N} =y_D v_H/(y_S v_S)$
is quite tiny for $v_S$ at weak-scale or higher. The smallness of $|V_{\ell N}|$ thus suppresses the massive eigenstate $N$'s effective couplings~\cite{Atre:2009rg} to weak gauge bosons as $|V_{\ell N}|^2\sim m_\nu/m_N$. 
Here we adopt this singlet scalar extension for its phenomenology simplicity: it is a low-cost extension to the SM that allows us to have an uncharged scalar around the weak scale that is only constrained by Higgs searches at colliders. Thus, we focus on the multi-lepton background analyses and search strategies. This scenario can also be realized in many UV-motivated models, like in supersymmetric models~\cite{Cao:2018iyk,Allahverdi:2014eca,Ma:2020mjz}, extra-gauge symmetry such as $U(1)_{B-L}$ extended scenarios~\cite{Deppisch:2018eth, Amrith:2018yfb}, etc., where additional constraints on the new physics sectors would apply. See~\cite{Cai:2017mow} for a recent theory review.

The extended scalar potential would contain $S$ self-interacting terms $V(S)$ and $H,S$ mixing terms. We would assume a small mixing term $\lambda v_S v_H \ \ll m_h^2$, so that $H$ and $S$ sectors minimize independently, and develop their vevs without qualitatively interfering with the electroweak breaking process. The mass matrix near the minimum can be written as
\be 
\begin{array}{c|cc}
& h & s \\
\hline
h & m^2_h &  {\lambda} v_H v_S \\
s & {\lambda} v_H v_S & m_s^2 \\
\end{array}.
\ee
where $h$ and $s$ are the neutral scalar modes near the minimum, with $(v_H+h)/\sqrt{2} \sim {\rm Re}{H^0}$. $m_h$ and $m_s$ represent the $h$ and $s$ masses from $H$ and $S$ potentials without the mixing term. Diagonalizing into the mass eigenstates, the scalars then mix by an angle $\alpha$,
\be  
\left( \begin{array}{c} h_1 \\ h_2\end{array} \right) = 
\left( \begin{array}{cc} 
\cos\alpha & -\sin \alpha \\ 
\sin\alpha & \cos\alpha 
\end{array} 
\right) 
\left( \begin{array}{c}h \\ s \end{array} \right).
\ee
where $h,s$ represent the Higgs doublet and singlet modes around their vevs. $h_1, h_2$ are the physical mass eigenstates. $h_1$ is dominated by $h$ and 
identified as the 125 GeV Higgs
boson. $h_2$ is singlet $s$ dominated, and it picks up a weakly charged $h$ component via mixing. $h_2$ is subject to diphoton resonance searches~\cite{Sirunyan:2018aui, Khachatryan:2015qba} and its mass range is less stringently constrained when the mixing angle is small,
\be 
\alpha = \frac{\lambda v_\mathit{H} v_S}{ |m_s^2 -m^2_h|},
\label{eq:sin_a}
\ee
In the small-mixing limit ($\lambda\rightarrow 0$), the denominator can be well approximated by $|m_s^2 -m^2_h|\sim |m_{h_1}^2 -m_{h_2}^2 +{\cal O}(\sin^2\alpha)|$ if the scalars are not mass-degenerate. Interestingly, if $h_2$ resides in the mass window $2 m_N<m_{h_2}< \sqrt{s} -m_Z$, production of $Z h_2$ is kinematically viable and $h_2\rightarrow NN$ decay can also contribute significantly to the signal. However, it would also require $m_{h_2}$ to be comparable to the Higgs boson's mass when the center-of-mass energy $\sqrt{s}$ is limited, particularly so if the $e^-e^+$ energy is just above the $Zh_1$ threshold at the Higgs factory. In this work, we focus on 240 GeV $e^+e^-$ center-of-mass energy, we assume $m_s$ to be generally outside this rather narrow window and only include $h_1$ production.

In case $m_s$ is much heavier than the weak scale, $S$ can be integrated out in the Lagrangian and the Higgs doublet can have an effective dim-5 $\bar{N}^C_RN_R H^\dagger H$ operator with a dimensionful coupling, like in $\nu$SMEFT setups~\cite{Caputo:2017pit, Barducci:2020icf}, where the only $h_1$ decays to $NN$ and any heavy scalars do not appear at low scale. Note at 240 GeV center-of-mass energy, $m_{s}$ isn't necessarily heavy for $h_2$ to be outside the collider's energy reach. 
An $h_2$ below $\sim$300 GeV causes very significant $h-s$ mixing in Eq.~(\ref{eq:sin_a}) unless $\lambda\ll 1$,
so when the mass window opens up for on-shell $h_2$ production, e.g. at 360 GeV or higher, resonant $h_2\rightarrow NN$ also becomes relevant, and $m_{h_2}$ alters both the total signal rate and the kinematics of the decay products~\cite{Gao:2019tio}. Since we are generally interested in resonant $NN$ production through these states, we will allow $S$ to have a weak scale mass. Our kinematic analysis and search strategies would be applicable to $h_2\rightarrow NN$ for future lepton collision runs at 360 GeV and 500 GeV center-of-mass energies.

The $h_1\rightarrow NN$ decay branching fraction is 
\be 
{\rm BR}_{h_1\rightarrow NN} = \Gamma_{h_1}^{-1}
\cdot  \frac{\left| \sin\alpha\cdot y_S \right|^2 m_{h_1}}{16\pi} \left(1-\frac{4m_N^2}{m_{h_1}^2} \right)^{3/2},
\ee
which assumes Majorana $N$ throughout this paper
\footnote{
For Dirac-type heavy neutrinos, the differences in decay fraction and kinematics require a separate analysis on cut efficiencies.
}, 
and taking the total Higgs boson width as $\Gamma_{h_1}\sim 4$ MeV. 
We would also assume $m_N < m_{h_1}/2$ so that the heavy neutrinos are produced on-shell. Note that current collider limits do not exclude a very light singlet-dominated scalar boson, $m_s \ll m_h$, which either indicates for a small $v_S$ (requiring $\lambda\ll 1$, plus an over-unity $y_S$ for a massive $N$), or requires $V(S)$ to be rather flat near its minimum with a large $v_S$. However, a very light scalar can not on-shell decay into $NN$ in our mass range of interest, thus we do not consider this scenario in this analysis.

\begin{figure}[h]
\centering
\includegraphics[scale=0.45]{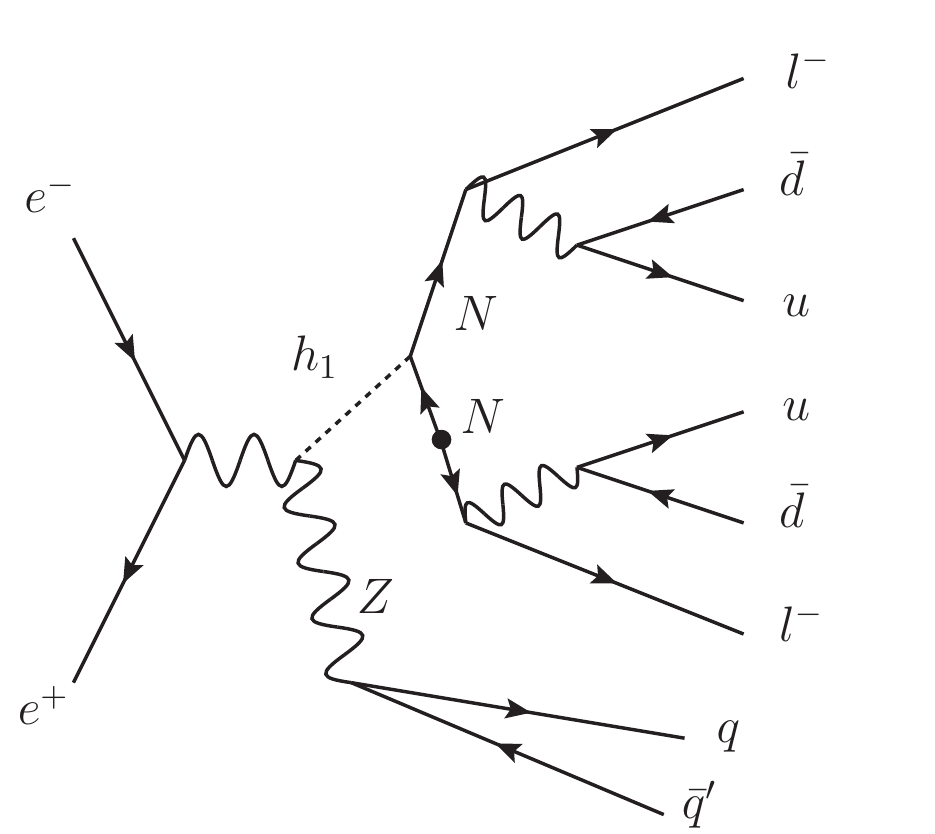}
\includegraphics[scale=0.45]{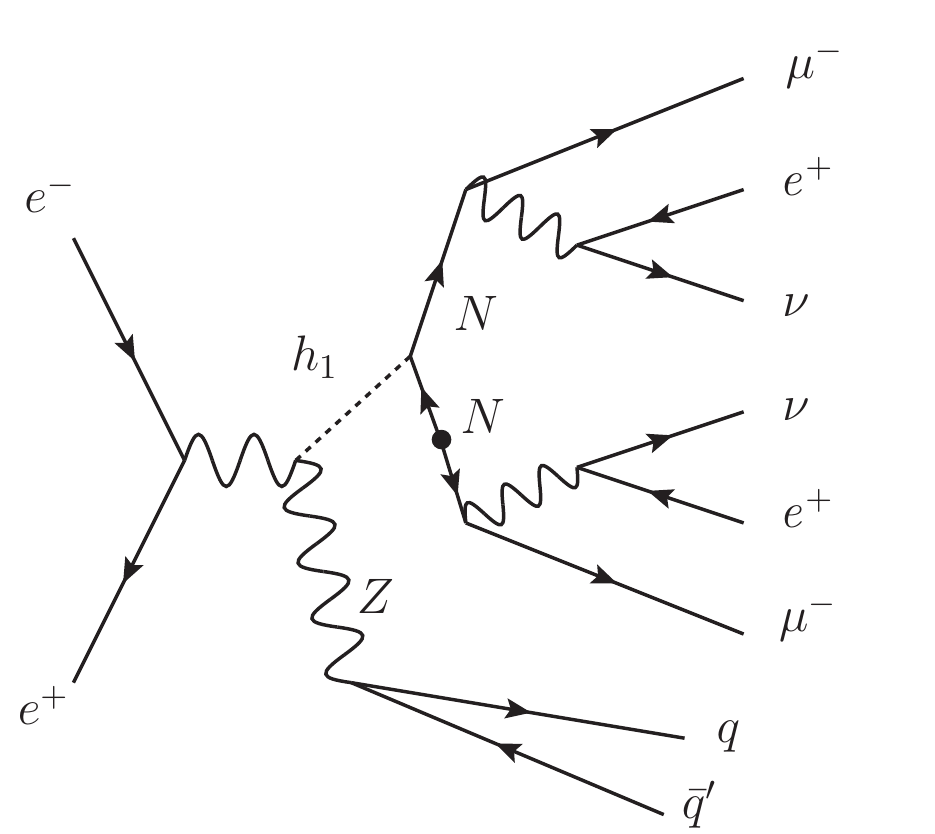}
\includegraphics[scale=0.45]{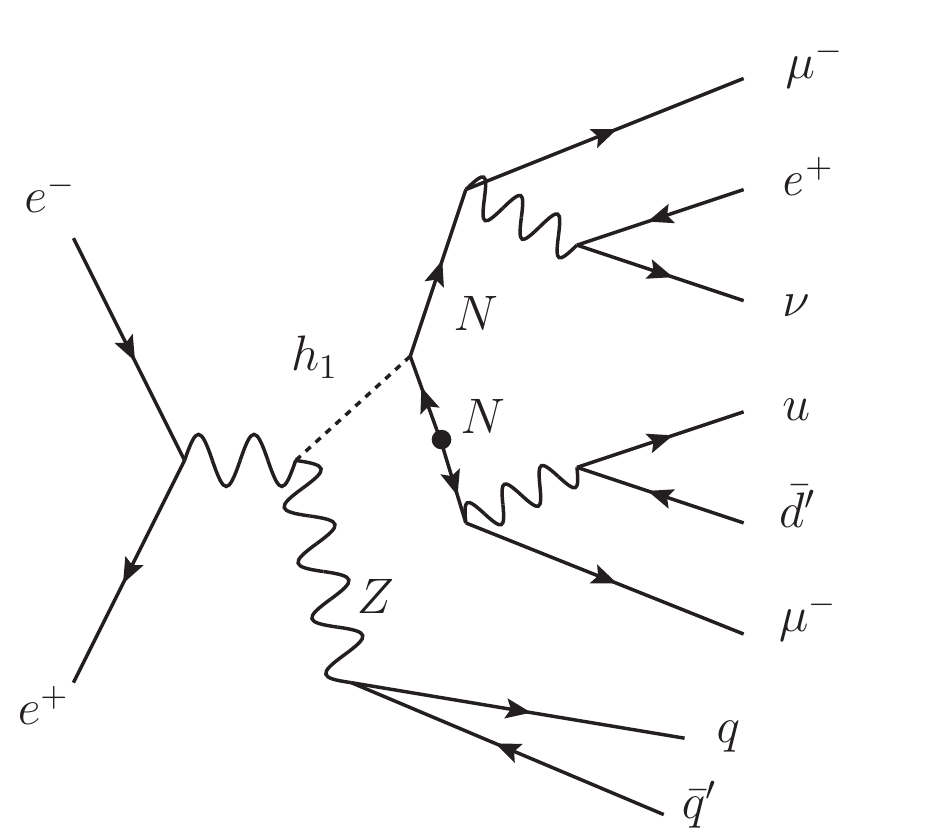}
\caption{Illustrative processes of SS dilepton production from semileptonic (upper left), leptonic (upper right) and mixed (lower) decays of the $NN$ system. }
\label{fig:feynman}
\end{figure}

Unlike Higgs production in $pp$ collision where gluon fusion dominates, the leading production channel in $e^-e^+$ collision is through the $s$-channel $Z$ with an associated final state $Z$-boson, as shown in Fig.~\ref{fig:feynman}. 
The $h_1\rightarrow NN$ system leads to the characteristic rare Higgs decay signals with SS lepton pair(s). 
Here we do not include the $h_1 \rightarrow N \nu$ channel~\cite{Das:2017rsu, Banerjee:2015gca, Das:2017zjc, Ding:2019tqq,Wang:2019xvx}, as this $h_1$ decay to $N\nu$ occurs via the left-handed component in $N$ and the effective coupling is proportional to $|V_{\ell N}|^2\sim m_\nu/m_N$. Consider the parameter range $1\gg |\sin \alpha|^2\gg m_\nu/m_N$, as will be shown later as relevant to collider sensitivity limits, the $h_1\rightarrow N N$ decay dominates over $h_1 \rightarrow N \nu$. Thus $h_1\rightarrow NN$ serves as a complementary search to $|V_{\ell N}|^2$-based channels in the presence of Higgs mixing with a BSM scalar.

The associated $Z$-boson can decay to a lepton or jet pair and their invariant mass reconstructs to $m_Z$. 
For signal selection, 
the decay mode of the $Z \to j j$ is favorable as it would:

(i) demand SM background to be also accompanied by one additional $Z$-boson; 
 
(ii) remove potential lepton number uncertainty from neutrinos in case LNV is required in the final state;
 
(iii) ensure the leptons are from the $NN$ decay.
 
The $h_1\rightarrow NN$ branching fraction is proportional to $\sin^2\alpha$. $N$ decays through its $\nu_L$ component and its partial width is suppressed by $|V_{\ell N}|^2$. $N\rightarrow \ell W^*$ is the dominant $N$ decay channel, its hadronic and leptonic $W^*\rightarrow jj, \ell\nu$ decays lead to `semileptonic' and `fully-leptonic' final states, see~\cite{Gao:2019tio} for branching fraction calculations. The $N\rightarrow \nu h_1^*, \nu Z^*$ channels are subleading and they require missing or wrong-sign leptons to form SS dileptons.

A major SM background consists of $\tau$-lepton or on-shell $W$ boson decays, where the $W\ell\nu$ vertex can couple to any lepton flavor. 
Same-sign and same-flavor dilepton
requires the presence of same-sign $\tau$ or $W$ pairs, and charge conservation would demand four $W^{(*)}\ell\nu$ vertices in a SM final state. 
Opposite-sign and same-flavor
dileptons arise from lepton pair production from neutral bosons, which can be vetoed by lepton charge cuts. 
Another background arises from wrong-sign leptons or missed leptons that can be controlled by stringent lepton cuts in event analysis. The following sections will discuss the backgrounds for each channel.

In a final state with three or more leptons, 
opposite-sign same-flavor (OSSF) lepton pairs should be vetoed to suppress the SM background, which also means at least one SSSF lepton pair will be selected. For some signal processes this requires $N$ to couple to more than one lepton flavor, and we assume $N$ couples equally to both $e$ and $\mu$ flavors in our analysis.

In this study, we consider heavy Majorana neutrinos for all three signal channels.
The two lepton channel has no missing energy in the final state and violates lepton number apparently. A Majorana $N$ is necessary to produced this signal.
For the three and four lepton channels, the final state has missing energy from neutrino(s) or antineutrino(s). 
Since the detectors at colliders cannot identify the differences between the neutrinos and antineutrinos. The final state with missing energy can be from LNV or LNC processes. Therefore, although our analyses is for heavy Majorana $N$ for all three signal channels, a Dirac $N$ would also produce the three and four lepton signal channels.
Once the heavy neutrinos are discovered from these two signal channels, more studies are needed to determine their Majorana or Dirac nature.

For convenience with collider analysis, we assume only one heavy neutrino is kinematically accessible in $h_1$ decay and it couples equally to $e$ and $\mu$, namely $|V_{eN}|=|V_{\mu N}|$ and $V_{\tau N}=0$. We ignore $V_{\tau N}$ in this study for two reasons: (i) the major SM background contamination are from channels with multiple $\tau$ leptons; (ii) the additional leptonic $\tau$ decay branching further reduces the signal rate.  
The collider sensitivity on $V_{\tau N}$ are thus expected to be less stringent.

We perform cut-and-count analyses on Monte-Carlo simulated signal and background events. 
For the signal simulation, the interaction terms Eq.~(\ref{eqn:Lag}) plus the SM Lagrangian are implemented in the FeynRules program~\cite{Alloul:2013bka} to create the model file in the Universal FeynRules Output (UFO) format~\cite{Degrande:2011ua}.
Events are generated by MadGraph5~\cite{Alwall:2014hca} and showered by Pythia8~\cite{Sjostrand:2006za,Sjostrand:2007gs} package. $\tau$ lepton decays are handled by TAUOLA~\cite{Jadach:1993hs} as is implemented in Pythia8. $e^-e^+$ detector simulation is performed by DELPHES~\cite{deFavereau:2013fsa} with CEPC parametrization~\cite{Chen:2017yel}. At event generation level we adopt jet cuts $\eta(j)<5.0,\, p_T(j)>20$ GeV and use relatively lenient lepton cuts $\eta(\ell)<2.5,\, p_T(\ell)>0.5$ GeV. Angular separation cuts $\Delta R(j,j),~\Delta R(\ell,\ell) > 0.3 $ are also implemented.
We note that at future lepton colliders, the $p_T$ triggers for jets could be lower than 20 GeV. However, for the multi-jet final state, the reconstruction quality and efficiency  will become worse when jets  $p_T < $ 10 GeV. Since our signals have harder $p_T$ of jets than the background (see for example, the $p_T(j)$ distributions in the FIG.~\ref{fig:Obs2l} ), a lower threshold will lead to more background and thus  we choose 20 GeV as benchmark analyses. 

\section{Two lepton channel}
\label{sect:semileptonic}

The $h_1\rightarrow NN\rightarrow \ell^\pm \ell^\pm+4j, \,\ell = e,\mu$ channel requires both $N$ decay to a charged lepton and two jets ($W^*\rightarrow\bar{q}q'$) and one of the $N$s decays as its own antiparticle. This final state has no missing energy and violates lepton number with $\Delta L =2$, and it is often considered the `smoking-gun' channel of the heavy Majorana neutrino search with explicit LNV.

Note in this channel $NN$ decays to four jets. Unlike being easily contaminated in $pp$ collision~\cite{Dib:2017iva,Antusch:2018bgr}, the much cleaner environment in $e^-e^+$ collision is largely free of fake leptons from soft jets. However, due to the small energy cap between the Higgs boson and heavy neutrino, these the four jets are relatively soft and can be difficult to fully reconstruct. With fewer jet counting, wrong-sign and unreconstructed leptons become possible backgrounds, in addition to intrinsic multiple-$\tau, W$ backgrounds. The relevant background channels are listed in Table~\ref{tab:2l}.

The background channels contain two or more $\tau$ leptons or light charged leptons $\ell=e,\mu$, plus $Z$ or $W^+W^-$ that yield dijet resonance near $Z$ mass. The particle signs are omitted and any even number of $\tau, \ell$ and $W$ entries must include equal number of opposite-sign particles, e.g. $4\tau$ denotes $2\tau^+2\tau^-$, and $2\ell 2W$ stands for $\ell^+\ell^-W^+W^-$, etc. While the SM backgrounds $\ell\ell$ are restricted to same-flavor, opposite sign lepton pairs, leptonic $W$ decay can provide one additional lepton and create a like-sign dilepton combination in case one lepton is un-detected. Such channels are marked with $^\dagger$ in the table.

The signal event contains one SS dilepton, two jets from $Z$ decay and a number of soft jets, plus very little missing energy. 
We select hadronic $Z$ decay by imposing charged lepton number $N(\ell)=2$ to avoid confusion between the leptons from $Z$ decay and those from $NN$ decay. With a jet transverse momentum requirement $p_T(j)>20$ GeV the soft jets from the $NN$ system are not always identifiable, especially when the jets are more collimated if $N$ is light and relatively boosted. Still, having at least one extra jet in addition to those from $Z$ can be effective in background rejection, so we consider the jet counting cut $N(j)\ge 3$.

Event selection includes the following cuts:

(i) exactly two leptons, $N(\ell)=2$ with $\, p_T(\ell)>5$ GeV; 

(ii) two leptons have the same sign;

(iii) veto $\tau$ leptons, $N(\tau)=0$;

(iv) at least three jets, $N(j) \ge 3$;

(v) small missing energy, $\met <15$ GeV.

\begin{figure}[h]
\centering
\includegraphics[width=4.0cm,height=3.0cm]{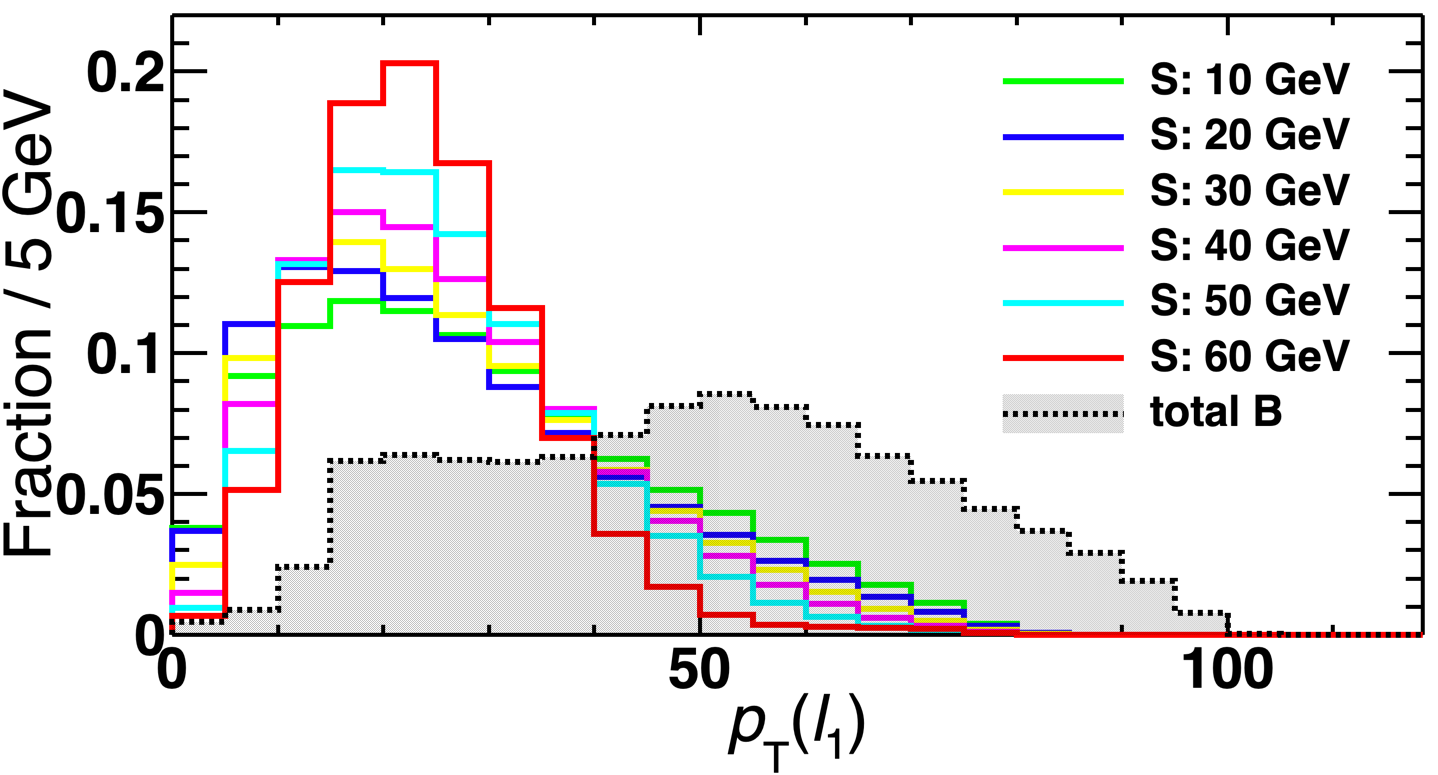}\,\,\,\,\,
\includegraphics[width=4.0cm,height=3.0cm]{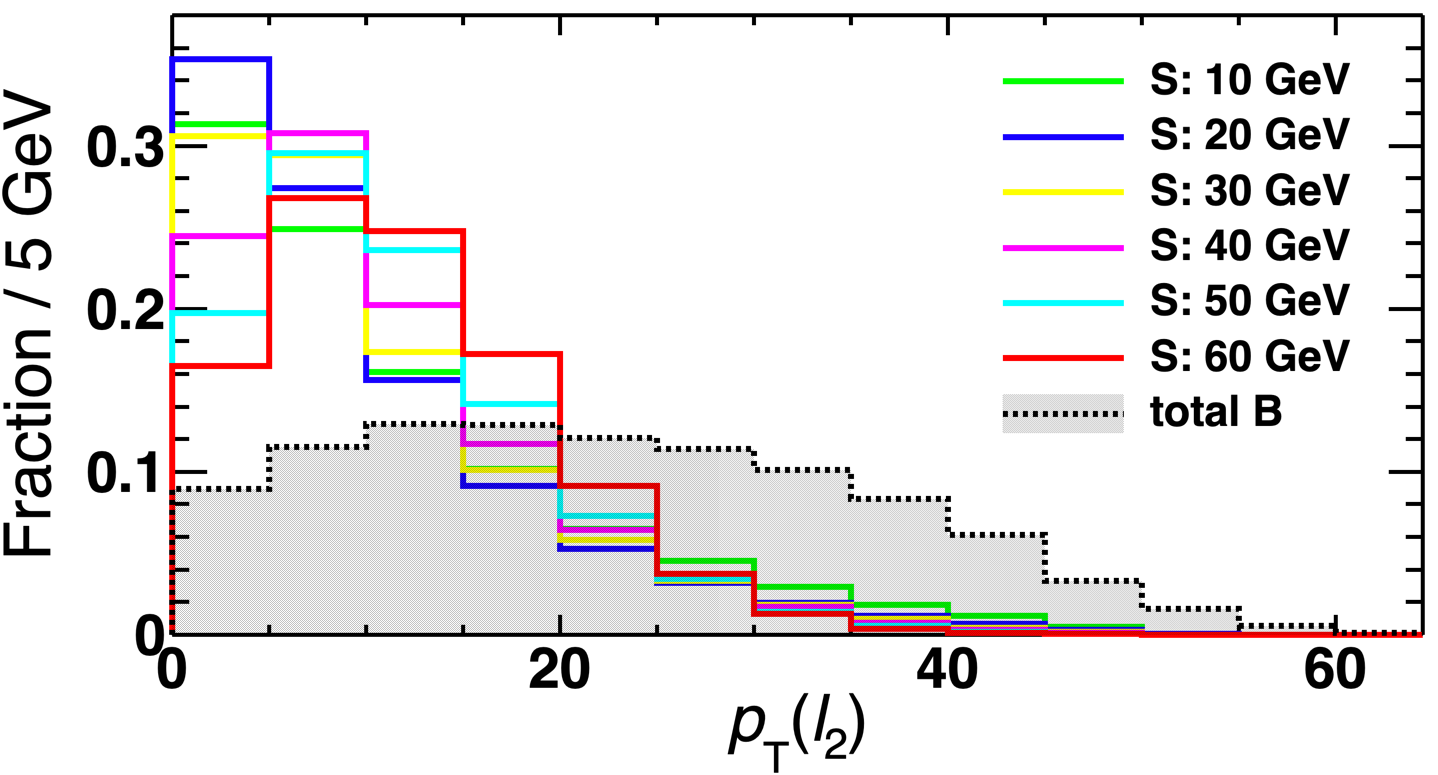}\,\,\,\,\,
\includegraphics[width=4.0cm,height=2.8cm]{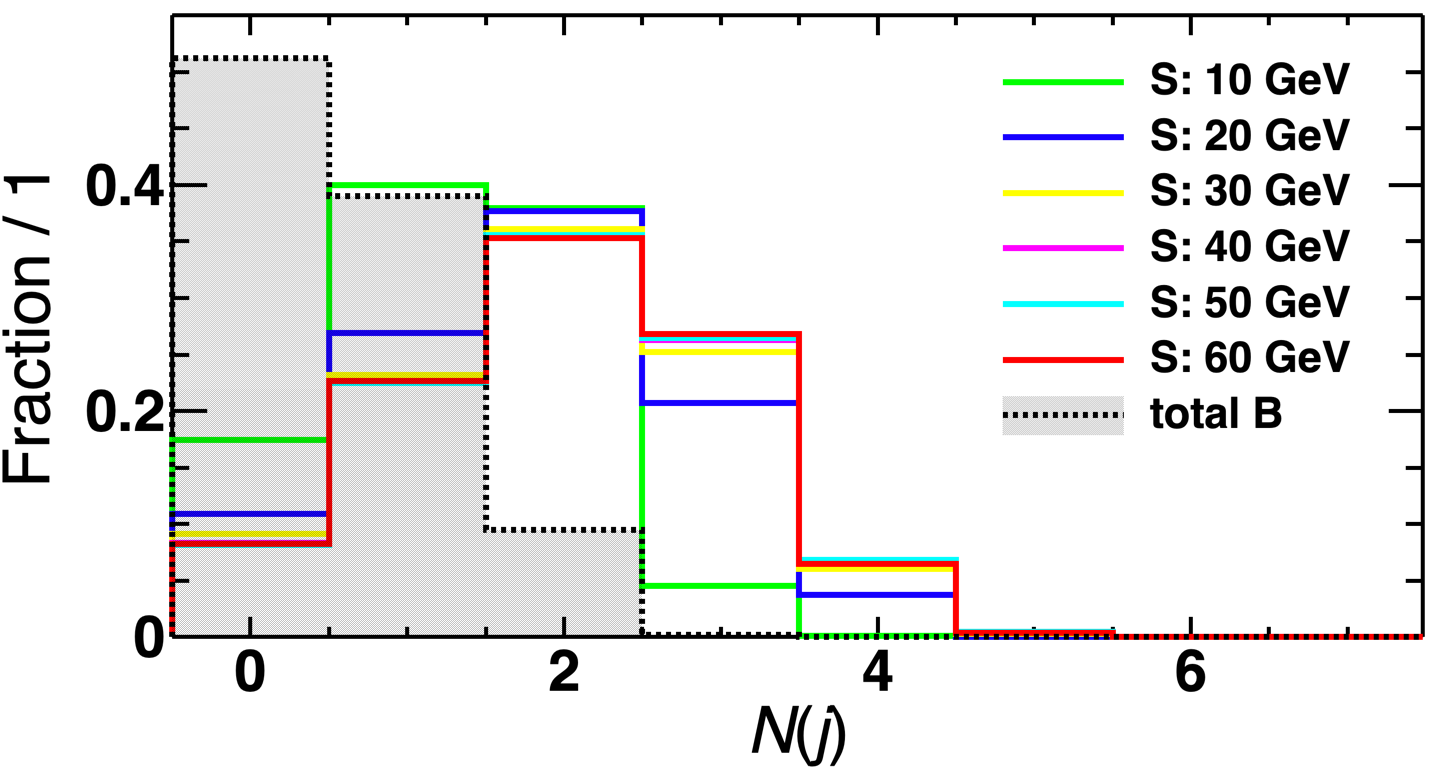}
\includegraphics[width=4.0cm,height=2.8cm]{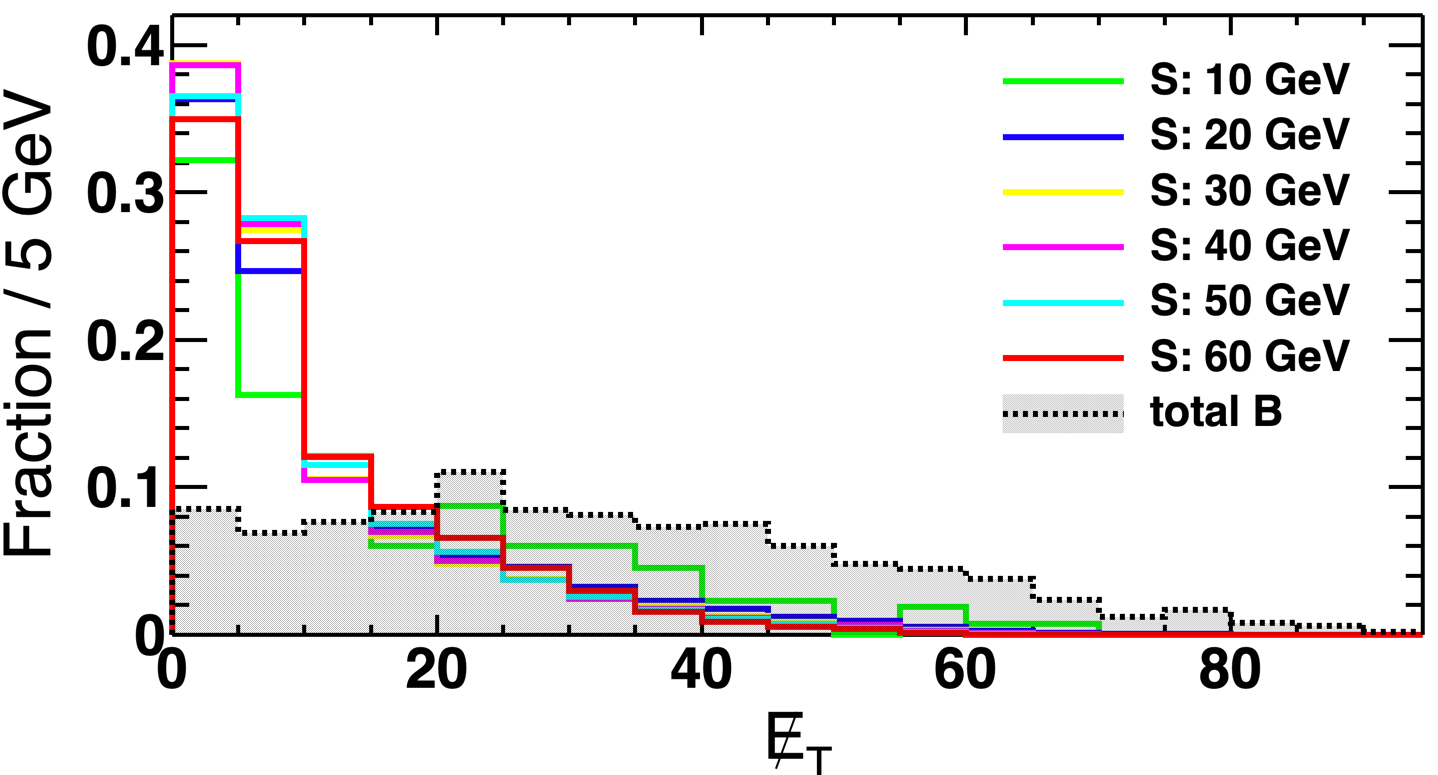}\,\,\,\,\,
\includegraphics[width=4.0cm,height=2.8cm]{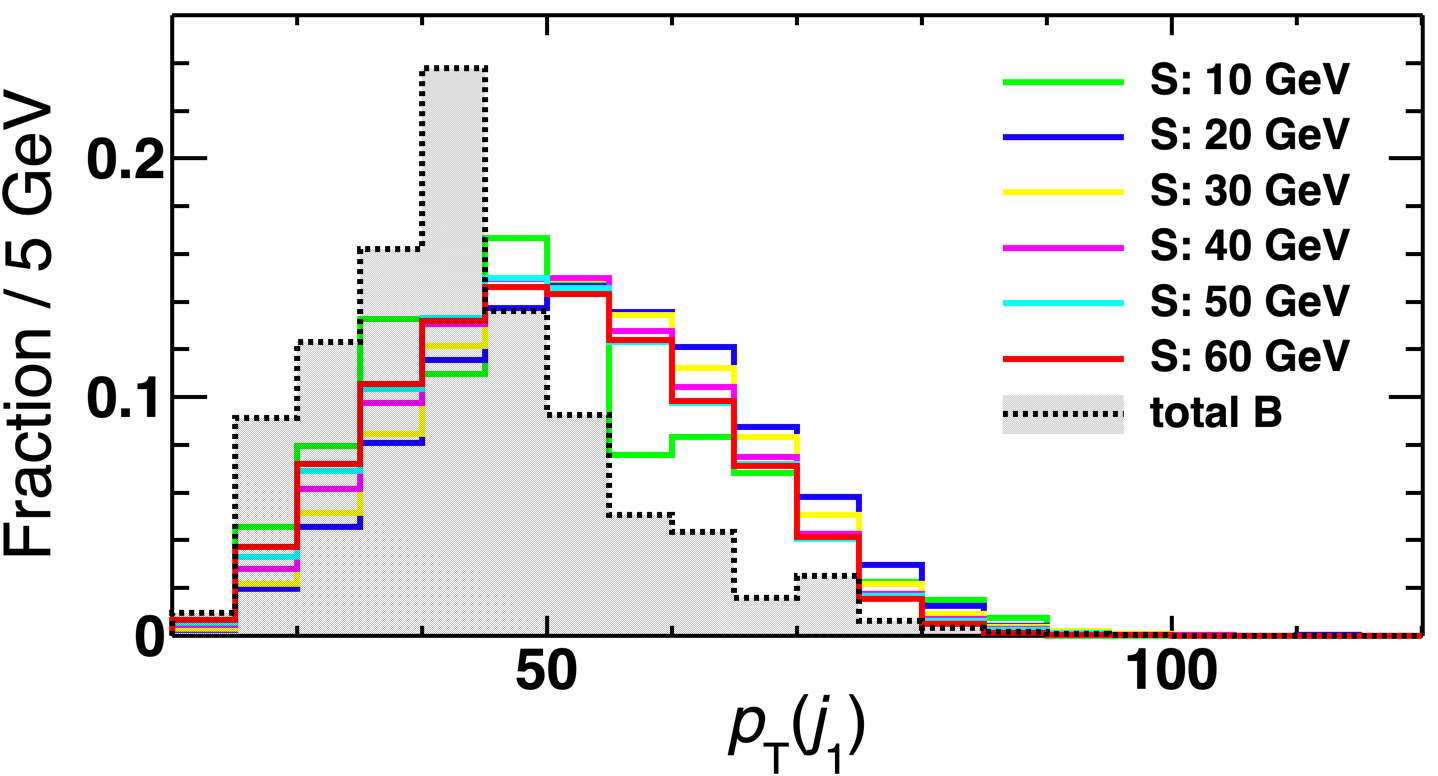}\,\,\,\,\,
\includegraphics[width=4.0cm,height=2.8cm]{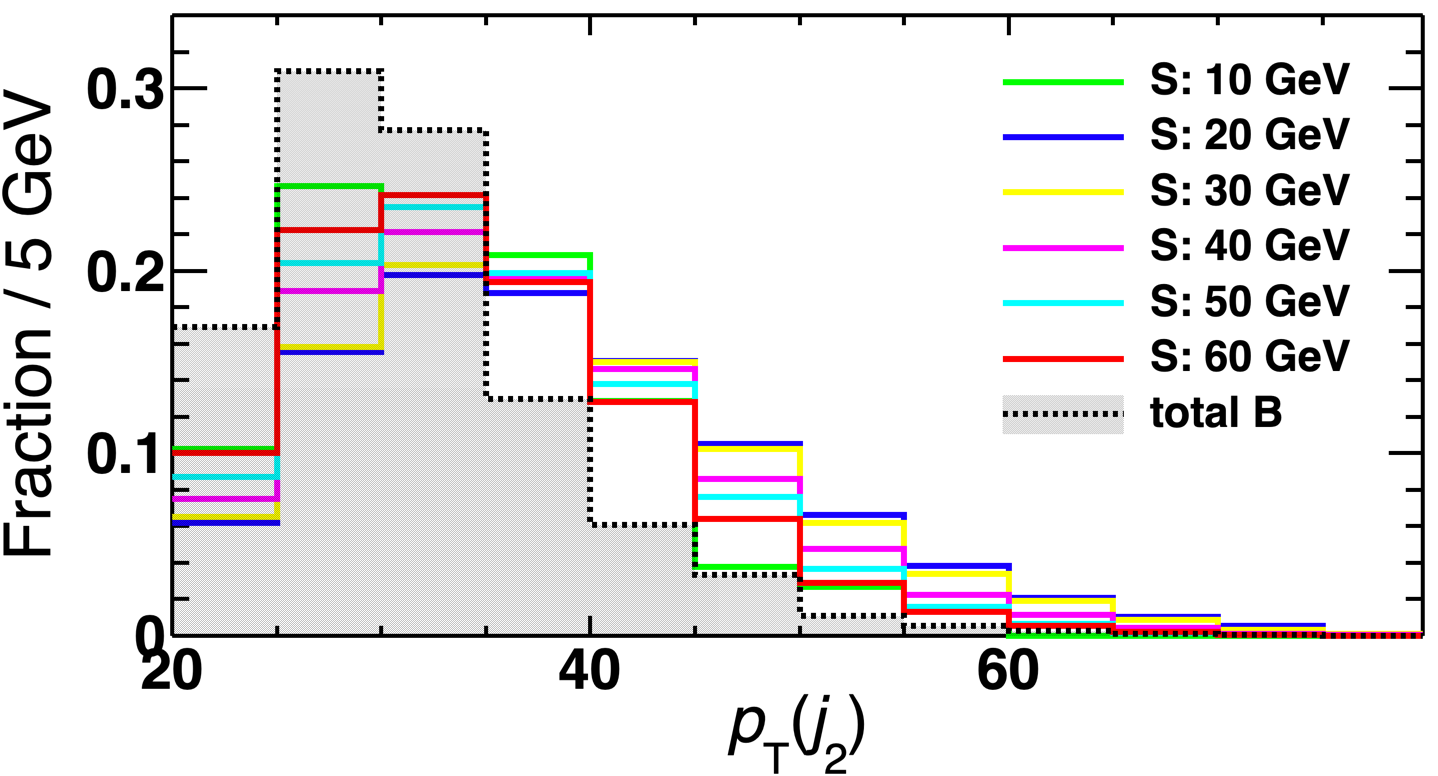}
\caption{Normalized distribution of selected kinematic variables to differentiate signals from the total SM background for the $2\ell$ channel. $p_T(\ell)$ panels are after selecting $N(l)=2$, $N(j)$ is after cuts(i-ii), $\met, p_T(j)$ panels are after cuts(i-iv).}
\label{fig:Obs2l}
\end{figure} 

The histograms of a few crucial kinetic observables for the signal with different $m_N$ and the total background are shown in Fig.~\ref{fig:Obs2l}. $p_T(\ell_1)$ and $p_T(\ell_2)$ correspond to the samples after requiring $N(\ell)=2$; $N(j)$ is after selecting cuts(i-ii); $\met$, $p_T(j_2), p_T(j_3)$ are after requiring cuts(i-iv). These distributions illustrate the effectiveness of our selection cuts. The $N(j)$ cut can be very effective in background rejection, and we select $N(j)\ge 3$ as it yields the best signal significance.

\begin{table}[h]
\centering
\begin{tabular}{c|c|cccc}
\hline
\hline
\multicolumn{2}{c|}{} & initial & cuts(i-ii) & cuts(iii-iv)& cuts(v)   \\
\hline
\multirow{6}{*}{Sig.} 
& 10 GeV  & $10^3$ & 6.3 & 0.29 & 0.18 \\
& 20 GeV  & $10^3$ & 35.9 & 8.8 & 6.4 \\
& 30 GeV  & $10^3$ & 72.3 & 22.6 & 17.5 \\
& 40 GeV  & $10^3$ & 97.2 & 32.5 & 25.3 \\
& 50 GeV  & $10^3$ & 112 & 37.4 & 28.8 \\
& 60 GeV  & $10^3$ & 121 & 40.5 & 30.2 \\
\hline
\multirow{8}{*}{Bkg.}
& $4\tau$    & $1.69\times10^4$   & 870  & $4.6\times10^{-2}$  & $7.7\times10^{-3}$ \\
& $^\dagger 2\tau Z$ & $6.80\times10^5$   & $2.91\times10^3$  & 4.6  & 0.93  \\
& $^\dagger 2\ell Z$  & $1.74\times10^6$   & $3.98\times10^3$  & - & -  \\
& $4\tau Z$ & 93.0   & 2.0  & 0.19  & $5.9\times10^{-2}$  \\
& $2\tau 2W$ & $4.42\times10^3$  & 63.6  & 0.92  & $8.2\times10^{-2}$  \\
& $^\dagger 2\ell 2\tau Z$ &  584  & 13.8 & 2.0 & 0.75 \\
& $^\dagger 4\ell Z$ & 862 &  16.5 & 2.2 & 2.1 \\
& $^\dagger 2\ell 2W$ & $2.74\times10^4$   & 639  & 11.7  & 1.2  \\
\hline
\hline
\end{tabular}
\caption{
The expected number of signal and background events for the $2\ell$ channel at future $e^-e^+$ collider with $\sqrt{s}=240$ GeV and 5.6 $\iab$ integrated luminosity. Signal rates assume a benchmark branching fraction BR$(h_1 \to NN) = 9.1\times10^{-4}$. Background channels marked with $^\dagger$ require wrong sign or missing leptons. Non-numeric dashes denote for event numbers below $10^{-3}$.
}
\label{tab:2l}
\end{table}

Since $t$-quark background is not a problem in $e^-e^+$ collision, $b$-jet veto is not included; $\tau$ veto is still helpful in removing multi-$\tau$ background events. The expected number of events at different cut stages for signal and background channels are listed in Table~\ref{tab:2l}. 
The benchmark branching fraction BR$(h_1 \to NN)$ is chosen to be $9.1\times10^{-4}$, so that the pre-cut (`initial') signal event rate is around $10^3$ and the signal cut efficiencies can be conveniently converted. The expected signal event number $N_s$ with selection cuts can be calculated from Eq.~(\ref{eqn:Ns}).

The major background includes $4\ell Z,~2\tau Z,~2\ell 2\tau Z$ and $2\ell 2W$ channels. The $4\ell Z$ and $2\ell 2\tau Z$ channels can fake a signal by missing two final state leptons with the same sign. In the $2\ell 2W$ channel, $W\rightarrow jj$ provides the required jets, and one missed lepton with the opposite sign can result in a fake signal event. As shown in Table~\ref{tab:2l}, these channels contribute more background events than the `intrinsic' $4\tau Z$ channel, and it shows the complication with selecting only one pair of same-sign leptons. 
The $2\tau Z$ background events are likely from one wrong-sign lepton.

\medskip
\section{Three Lepton Channel}
\label{sect:mixed}

When one $N$ decays leptonically and the other $N$ decays semileptonically, the three leptons in the final state $Zh_1\rightarrow \ell^\pm \ell^\pm \ell +4j +\met$ may contain one SSSF dilepton combination. Compared to the 2$\ell$ channel, the SSSF dilepton signal occurs with both lepton number violating ($\Delta L =2$) and conserving ($\Delta L = 0$) decays of $NN$. The $\Delta L = 2$ process is shown in Fig.~\ref{fig:feynman}.
When $N$ couples to at least two lepton flavors (e.g. to both $e$ and $\mu$), The $\Delta L = 0$ process can also obtain the same-sign lepton from the secondary leptonic $W^*\rightarrow \ell\nu$ vertex instead of the primary $N\rightarrow \ell W$ vertex, 
and thus this signal does not confirm that $N$ is Majorana-type.

Similar to the $2\ell$ channel, the SM background includes channels with multiple $\tau$ and $W$ bosons.
The OSSF lepton pair (e.g. $e^\pm e^\mp$) should be vetoed to suppress the SM background.

We select signal events with the following cuts:

(i) exactly three leptons $N(\ell) = 3$ with $p_T \ge 5$ GeV;

(ii) veto OSSF lepton pairs;

(iii) veto $\tau$ leptons, $N(\tau)=0$;

(iv) at least two jets, $N(j) \ge 2$.

\begin{figure}[h]
\centering
\includegraphics[width=4.0cm,height=2.8cm]{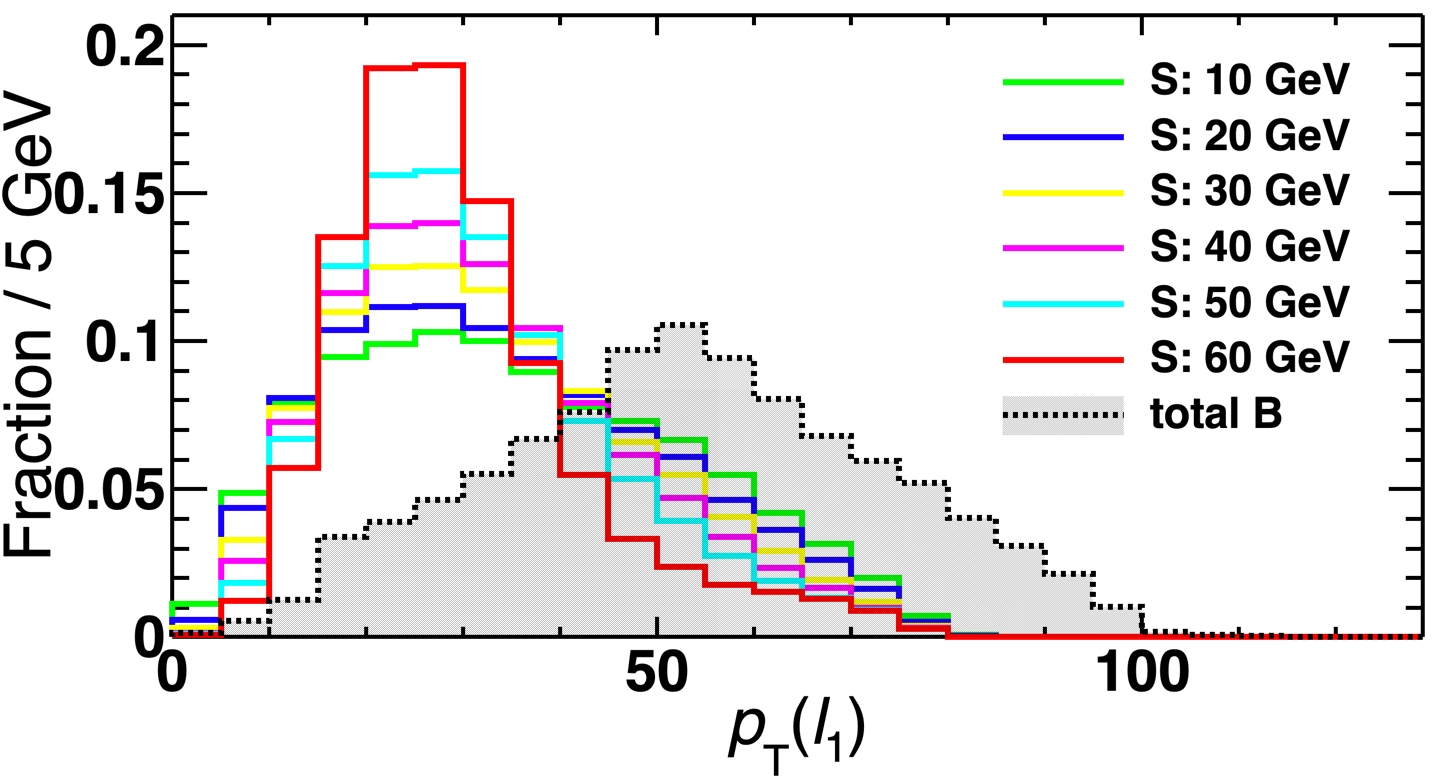}\,\,\,\,\,
\includegraphics[width=4.0cm,height=2.8cm]{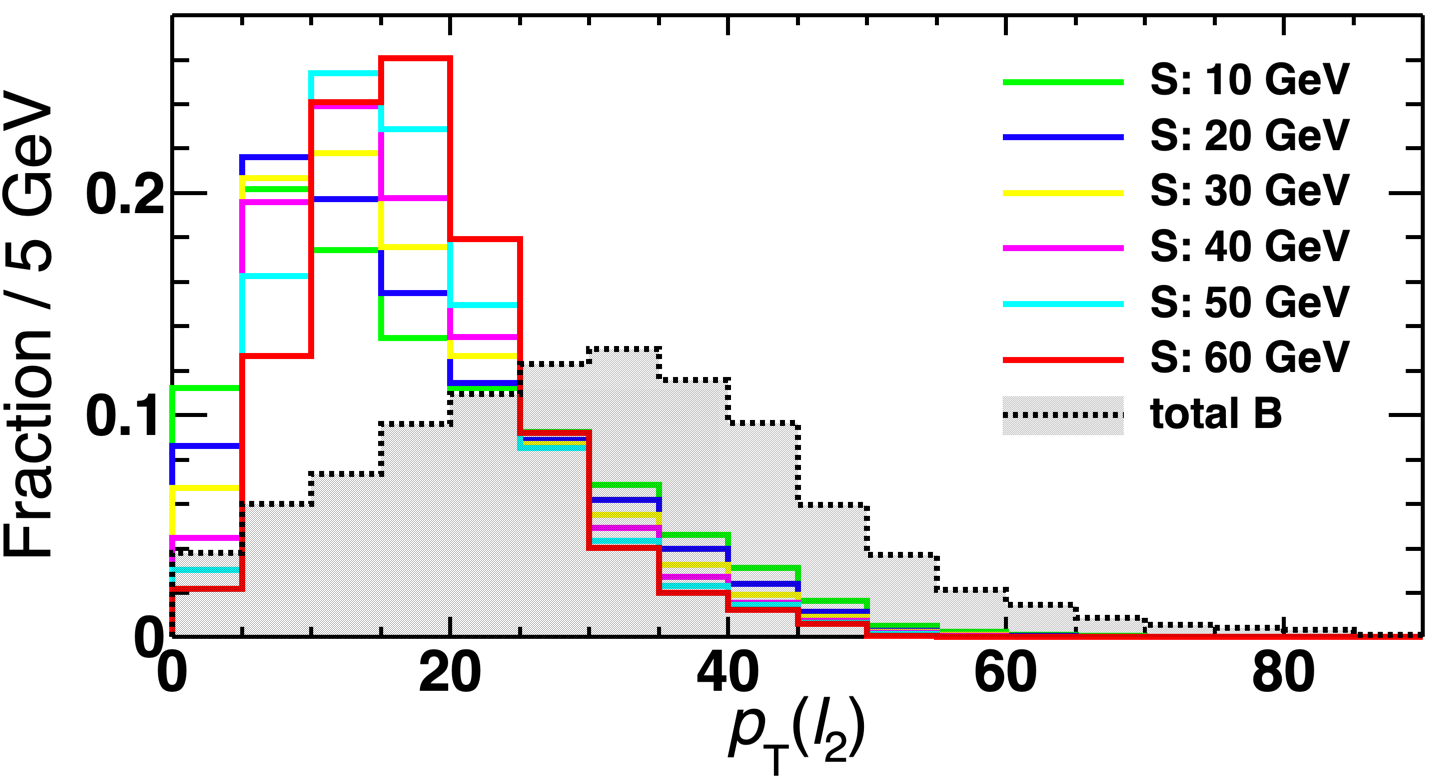}\,\,\,\,\,
\includegraphics[width=4.0cm,height=2.8cm]{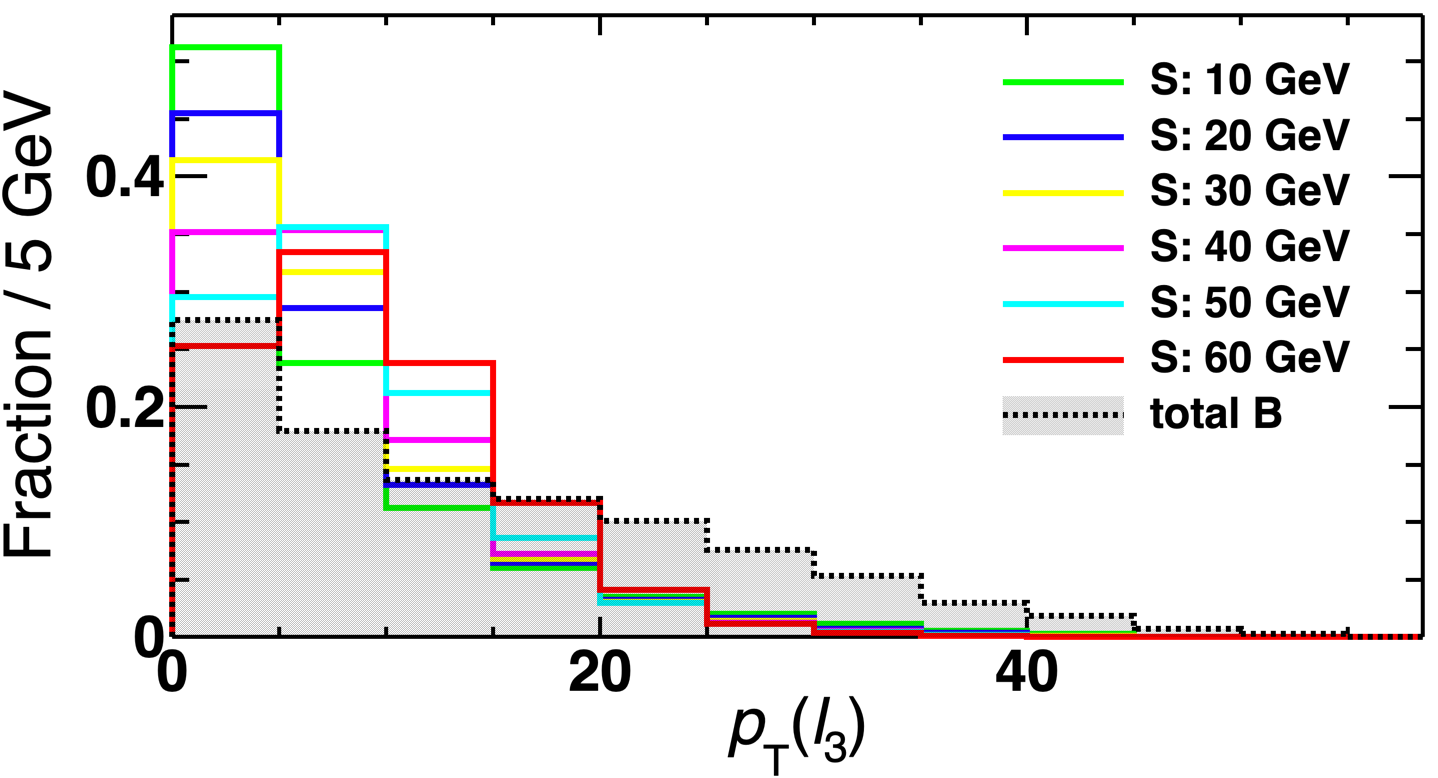}
\includegraphics[width=4.0cm,height=2.8cm]{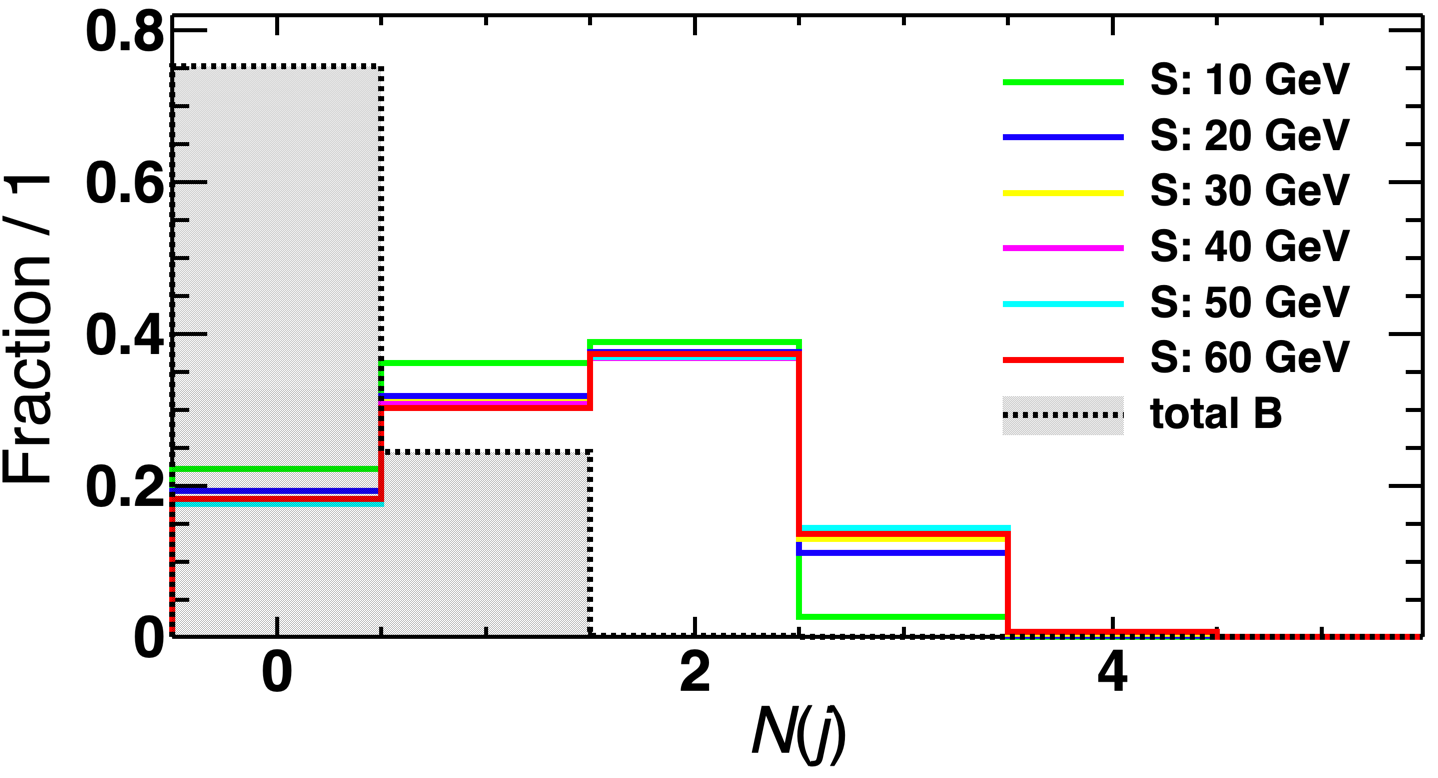}\,\,\,\,\,
\includegraphics[width=4.0cm,height=2.8cm]{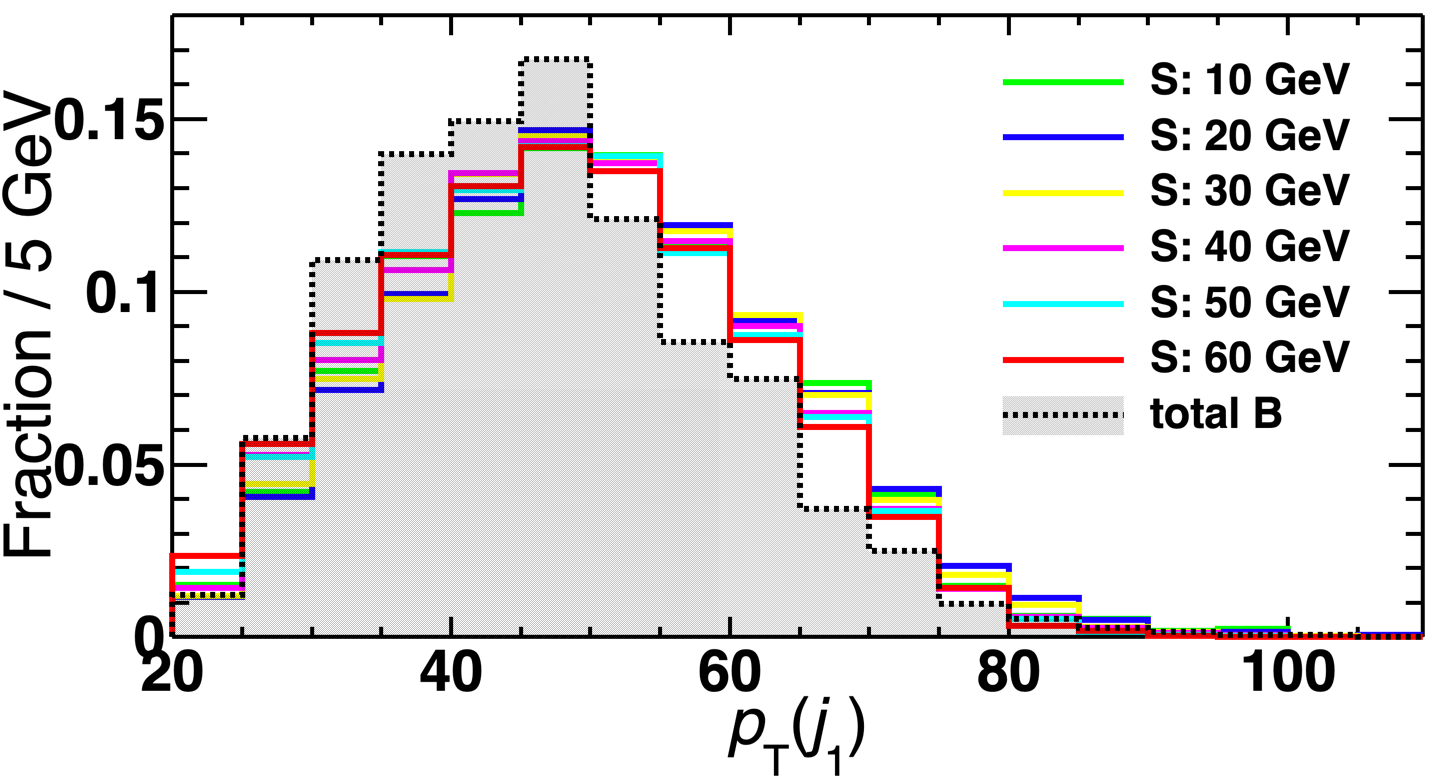}\,\,\,\,\,
\includegraphics[width=4.0cm,height=2.8cm]{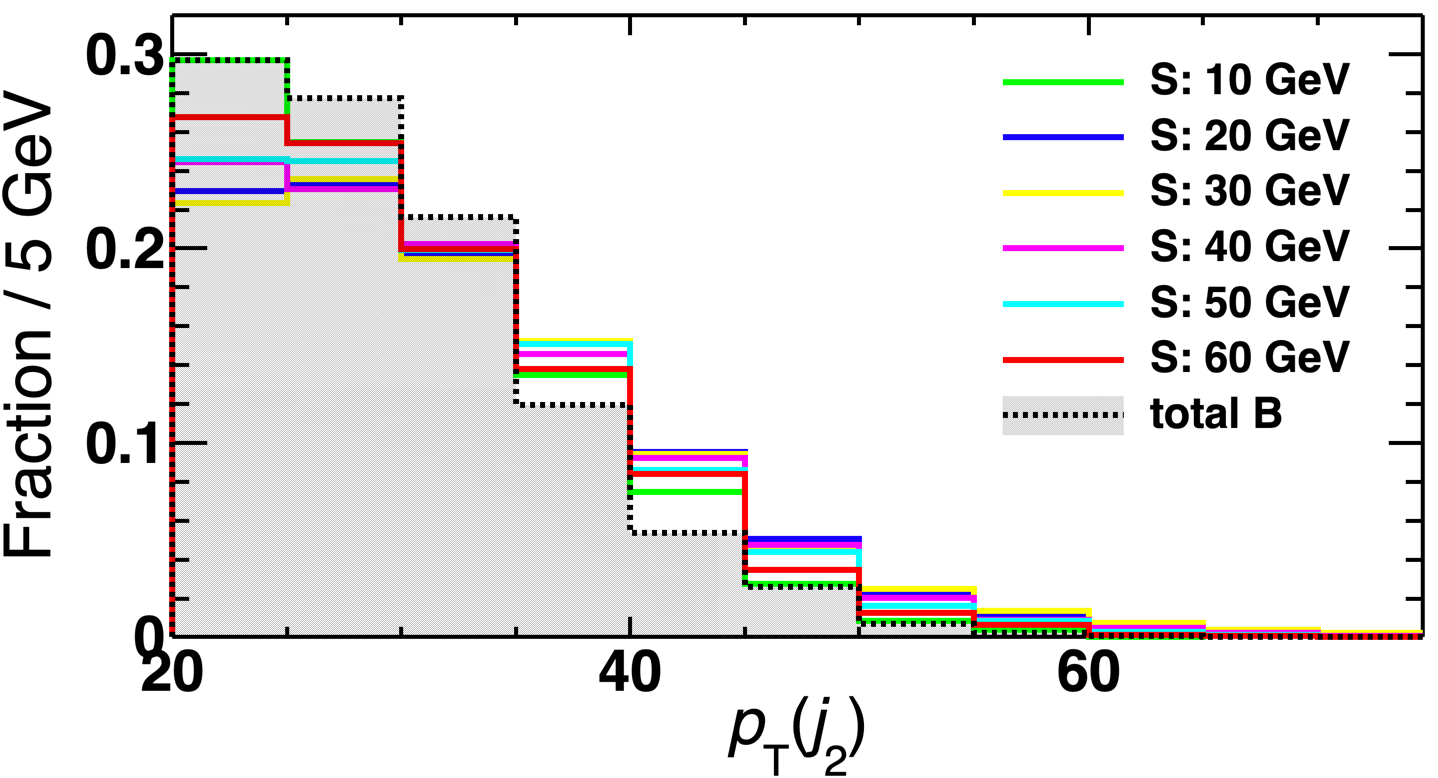}
\caption{
Normalized distribution of selected kinematic variables for the $3\ell$ channel. 
$p_T(\ell)$ panels are after selecting $N(l)=3$, $N(j)$ is after cuts(i-ii), $p_T(j)$ panels are after cuts(i-iv).
}
\label{fig:Obs3l}
\end{figure} 

The distributions of selected kinetic variables in the $3\ell$ channel are shown in Fig.~\ref{fig:Obs3l}. $p_T(\ell_1)$, $p_T(\ell_2)$, $p_T(\ell_3)$ correspond to the samples after requiring $N(\ell)=3$; $N(j)$ is after selecting cuts(i-ii); $p_T(j_1), p_T(j_2)$ are after requiring cuts(i-iv). $N(j)$ cut is selected to optimize the signal significance.

\begin{table}[h]
\centering
\begin{tabular}{c|c|cccc}
\hline
\hline
\multicolumn{2}{c|}{} & initial & cuts(i) & cuts(ii)& cuts(iii-iv)   \\
\hline
\multirow{6}{*}{Sig.} 
& 10 GeV  & $10^3$ & 27.9 & 5.6 & 2.3 \\
& 20 GeV  & $10^3$ & 62.7 & 13.6 & 6.6 \\
& 30 GeV  & $10^3$ & 85.8 & 19.9 & 10.0 \\
& 40 GeV  & $10^3$ & 102 & 24.9 & 12.7 \\
& 50 GeV  & $10^3$ & 112 & 27.3 & 14.1 \\
& 60 GeV  & $10^3$ & 115 & 28.2 & 14.4 \\
\hline
\multirow{8}{*}{Bkg.}
& $4\tau$    & $1.69\times10^4$   & 614  & 155  & $3.8\times10^{-2}$ \\
& $^\dagger 2\tau Z$ & $6.80\times10^5$   & $1.30\times10^4$  & 350  & -  \\
& $^\dagger 2\ell Z$  & $1.74\times10^6$   & $5.03\times10^4$  & 121 & -  \\
& $4\tau Z$ & 93.0   & 2.1  & 0.25  & $7.3\times10^{-2}$  \\
& $2\tau 2W$ & $4.42\times10^3$  & 27.8  & 6.9  & 0.72  \\
& $^\dagger 2\ell 2\tau Z$ &  584  & 46.5 & 1.1 & 0.44 \\
& $^\dagger 4\ell Z$ & 862 &  132 & 0.27 & $1.4\times10^{-2}$ \\
& $^\dagger 2\ell 2W$ & $2.74\times10^4$ & $1.30\times10^3$ & 37.8  & $5.0\times10^{-2}$  \\
\hline
\hline
\end{tabular}
\caption{
Similar to Table~\ref{tab:2l} but for the 3$\ell$ channel.
Background channels with $^\dagger$ require missing leptons.
}
\label{tab:3l}
\end{table}

Table~\ref{tab:3l} lists the expected number of events at different cut stages for signal with different $N$ masses and background channels.
In clear contrast to the $2\ell$ channel, the surviving backgrounds are $2\tau 2W$ and $2\tau 2l Z$ channels. The combination of $N(j)$ cut and increased lepton number cut effectively remove the background from leptonic $W$ decays, which leads to a smaller total background event rate. Because of both a lower leptonic $N$ decay branching fraction and fewer jets from $NN$ decays,
the signal event rate is also lower compared to the $2\ell$ channel.

\begin{figure}[h]
\centering
\includegraphics[scale=0.47]{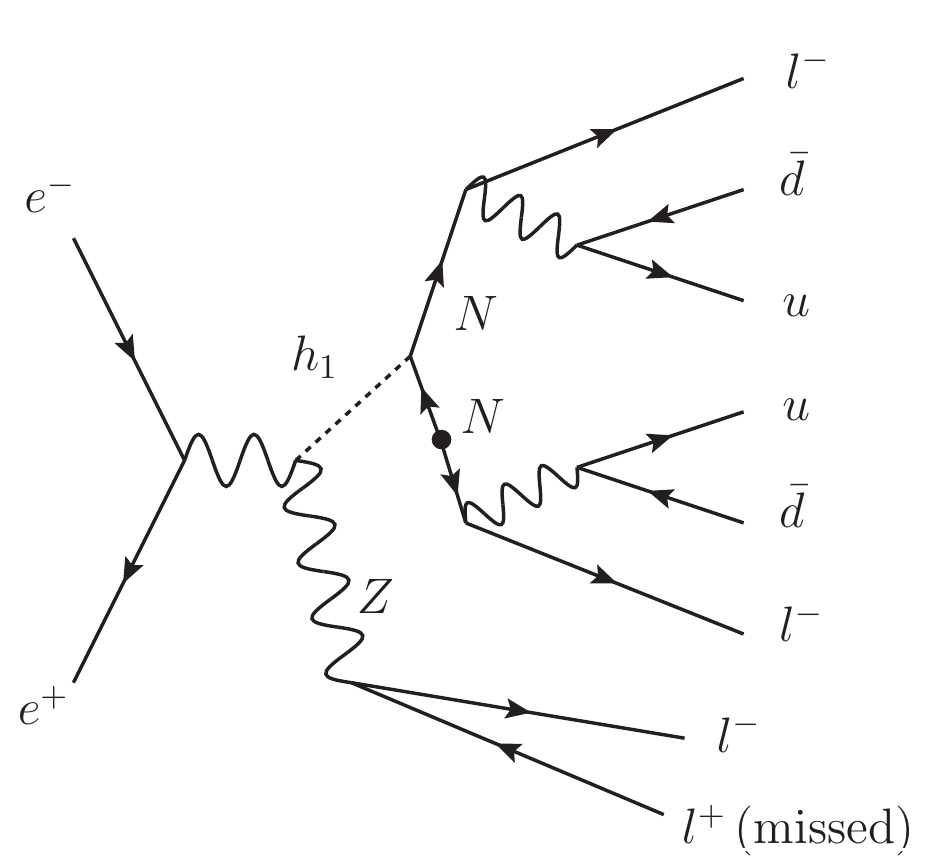}
\caption{SS trilepton emerges when $Z$ decay leptonically and the oppose-sign lepton misses detection.}
\label{fig:trilepton}
\end{figure} 

Interestingly, selecting three leptons will pick up a same-sign {\it trilepton} final state of $\ell^\pm\ell^\pm\ell^\pm$ that derives from leptonic $Z$ decay, as shown in Fig.~\ref{fig:trilepton}. The jets from semileptonic $NN$ decay satisfy the jet cuts, $\ell^\pm\ell^\pm\ell^\pm$ emerges once the opposite-sign $\ell^\mp$ misses detection. 
Among $\ell^\pm\ell^\pm\ell^\pm$ events, a fraction of 
$\frac{1}{2}(x_e^2+x_\mu^2)/\sum_{\ell, \ell^\prime} x_\ell x_{\ell^\prime}$ 
have the same flavor for all three leptons (i.e. SSSF trilepton $e^\pm e^\pm e^\pm$ and $\mu^\pm\mu^\pm\mu^\pm$), where $x_{\ell=e,\mu}$ denotes the branching fraction of $N$ decay into each lepton flavor that is proportional to $|V_{\ell N}|^2$. At our choice of $x_e:x_\mu =1:1$, one quarter of the SS trileptons have the same flavor.

At $m_N = 60$ GeV, this same charge trilepton final state $\ell^\pm\ell^\pm\ell^\pm$ is 7.6\% of the $\ell^\pm\ell^\pm\ell$ signal events after selecting cuts(i-ii). 
In comparison, SM events would need at least three missed or wrong-sign leptons to fake such a process. The SSSF trilepton background rate is found to be about 0.1\% of the original $3\ell$ background event rate after cuts (i-ii). If capped by luminosity limits, SSSF trilepton signal doesn't necessarily yield stronger sensitivity as its expected signal event rate is small.

\medskip
\section{Four Lepton Channel}
\label{sect:fullyleptonic}

The fully leptonic $NN$ decay leads to four charged leptons. When $N$ couples to both the first and second lepton generations, two SSSF dileptons $e^\pm e^\pm\mu^\mp \mu^\mp$ can emerge, and the two pairs must be in different flavors to avoid the OSSF dilepton pairs. Due to the presence of (anti)neutrinos, this final state does not guarantee LNV, and receives contribution from both LNV and non-LNV decays of $N$. 
Therefore, similar to Sec.~\ref{sect:mixed}, this signal does not require $N$ must be Majorana.

The fully-leptonic branching fraction is lower than the semileptonic branching fraction due to the smaller leptonic $W^*\rightarrow l\nu$ branching compared to the hadronic $W^*\rightarrow jj$ branching, plus the requirement that the two dileptons must differ in flavor. Having two SSSF dileptons can provide {\it major} reduction on backgrounds, and it is shown that the SM background can be below single-event level in $pp$ collision~\cite{Gao:2019tio}. At $e^-e^+$ collision, even lower background is expected, and it would be interesting to investigate at what luminosity level the background becomes relevant.

Similar to the semileptonic case, the SM background arise from multiple $\tau, W$ channels with one associated $Z$-boson. We consider the following selection cuts in event analysis:

(i) exactly four leptons, $N(\ell) = 4$ with $p_T(\ell)\ge 5$ GeV;

(ii) exactly two electrons with the same charges; exactly two muons with the same charges; electrons and muons have opposite charges; i.e. exactly $e^\pm e^\pm \mu^\mp \mu^\mp$ lepton pairs;

(iii) veto $\tau$ leptons, $N(\tau)=0$;

(iv) at least one jet, $N(j) \ge 1$.

\begin{figure}[h]
\centering
\includegraphics[width=4.0cm,height=2.8cm]{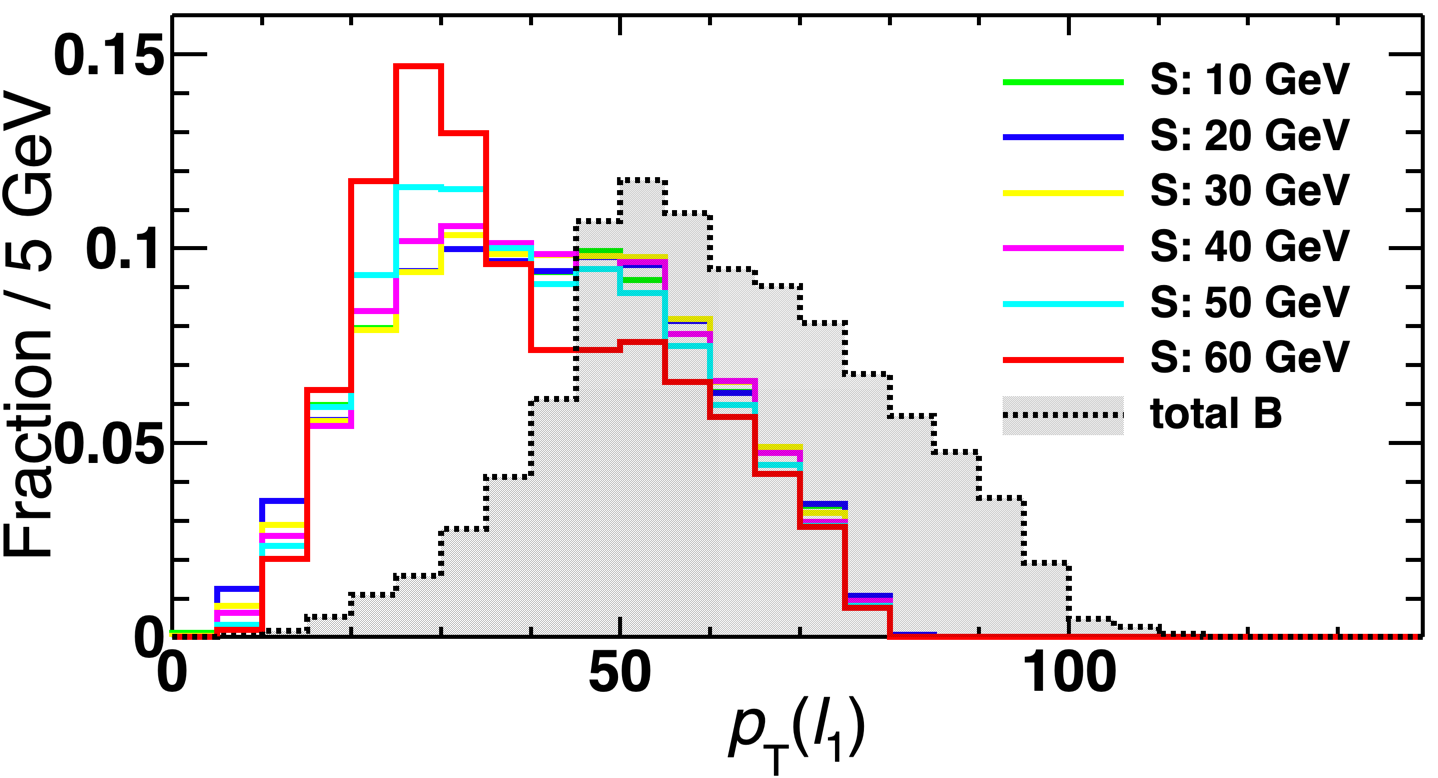}\,\,\,\,\,
\includegraphics[width=4.0cm,height=2.8cm]{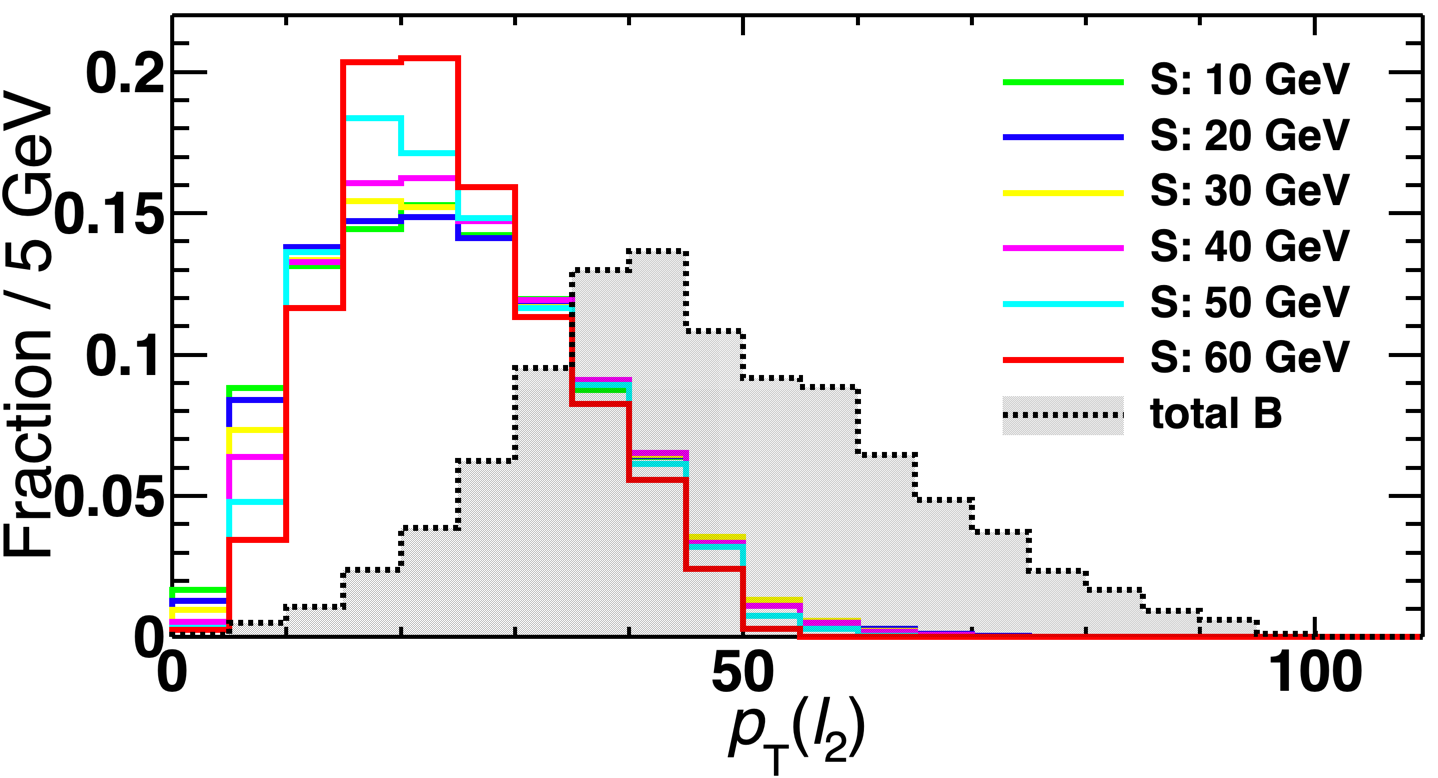}\,\,\,\,\,
\includegraphics[width=4.0cm,height=2.8cm]{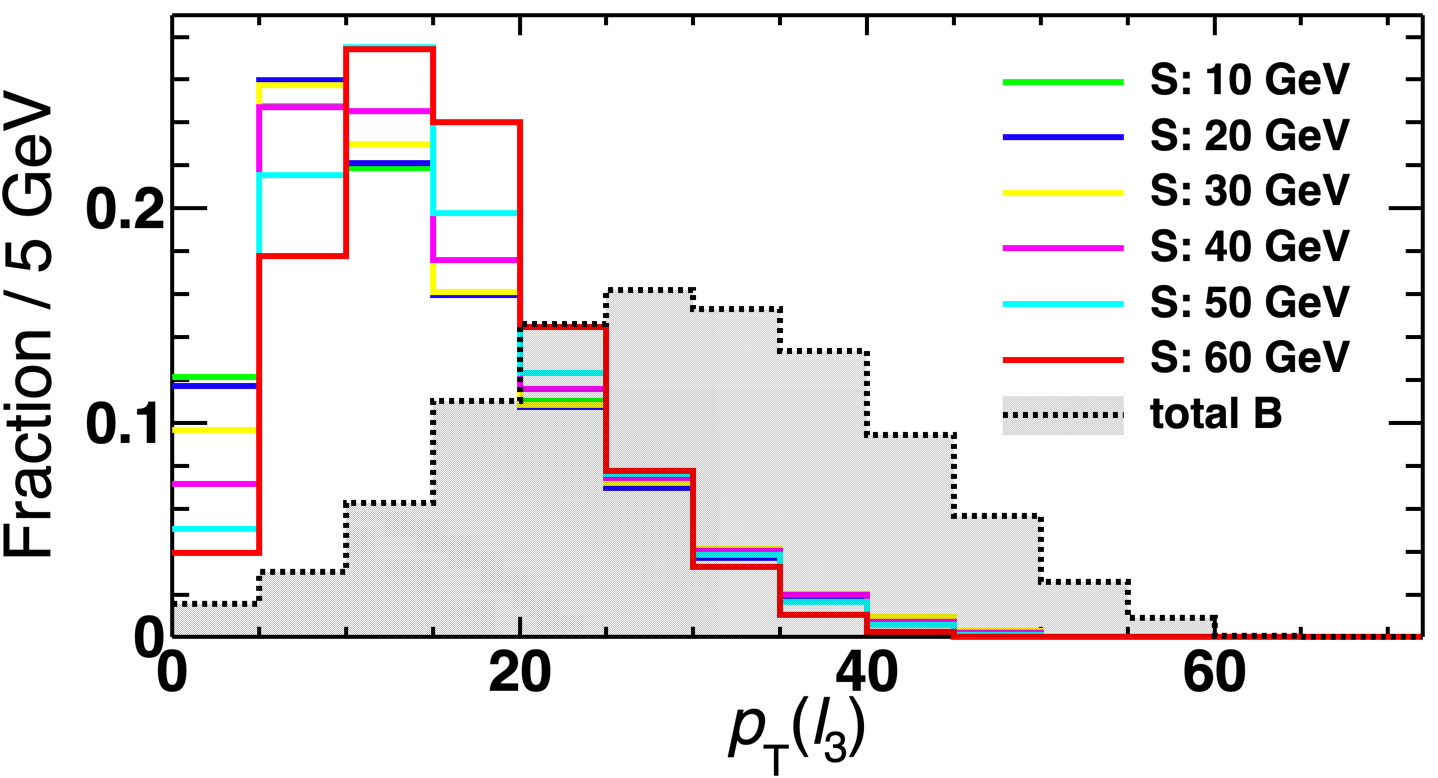}
\includegraphics[width=4.0cm,height=2.8cm]{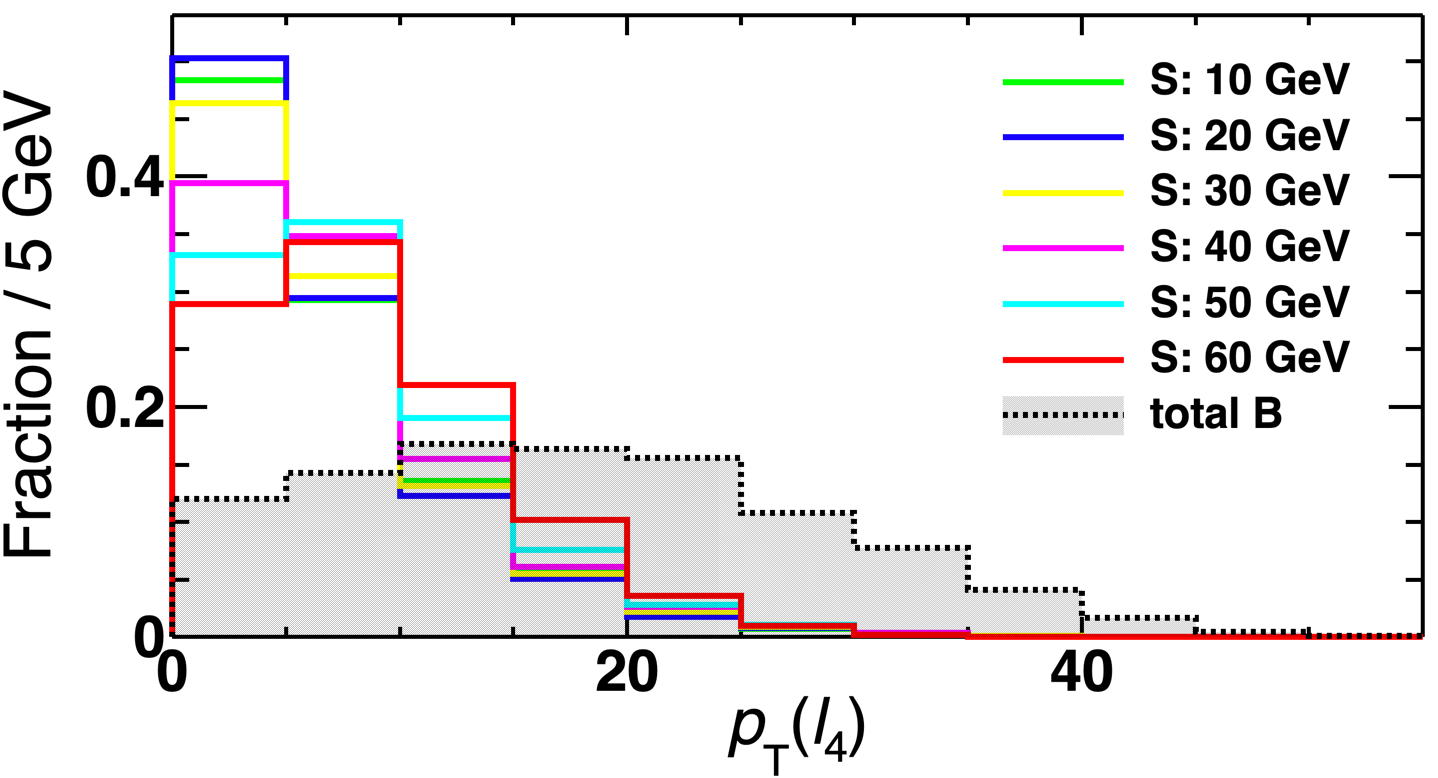}\,\,\,\,\,
\includegraphics[width=4.0cm,height=2.8cm]{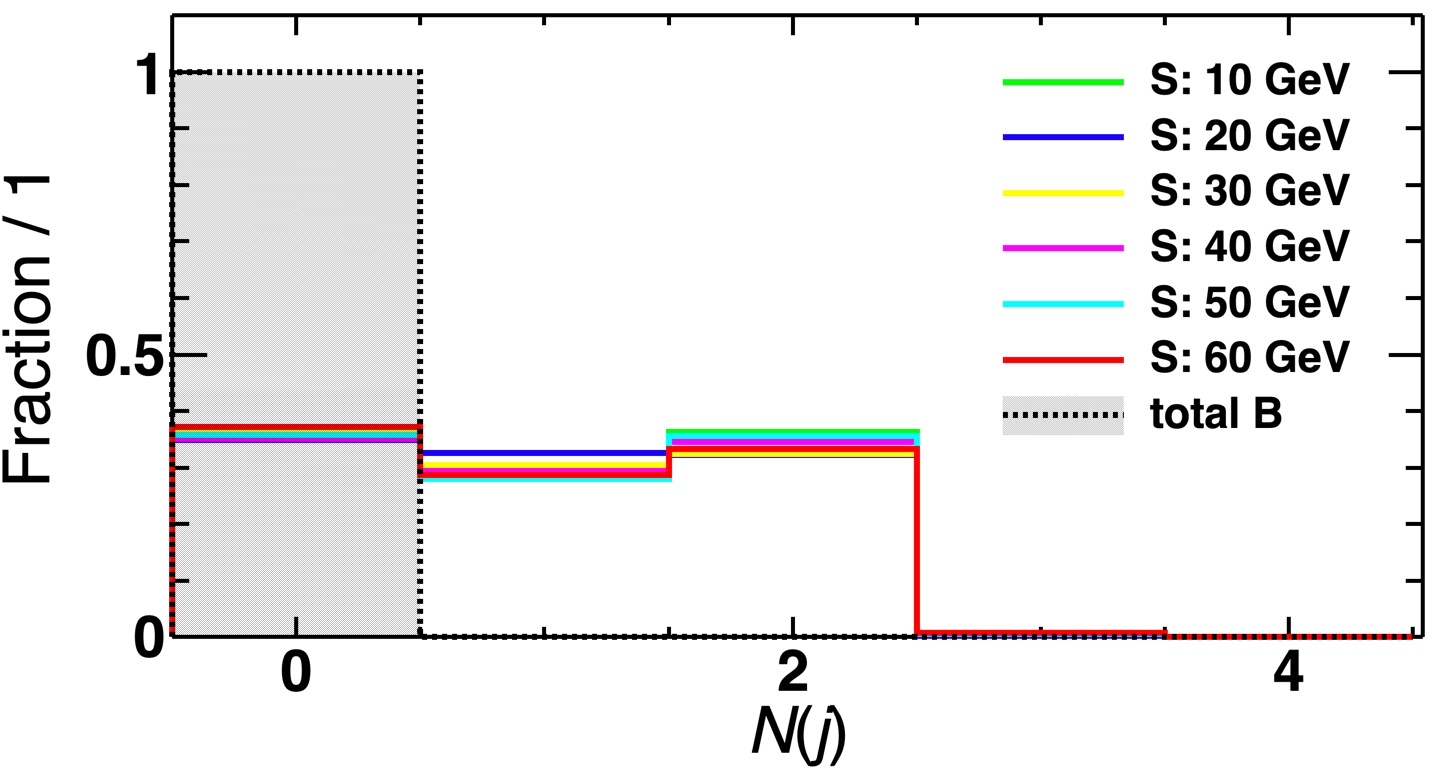}\,\,\,\,\,
\includegraphics[width=4.0cm,height=2.8cm]{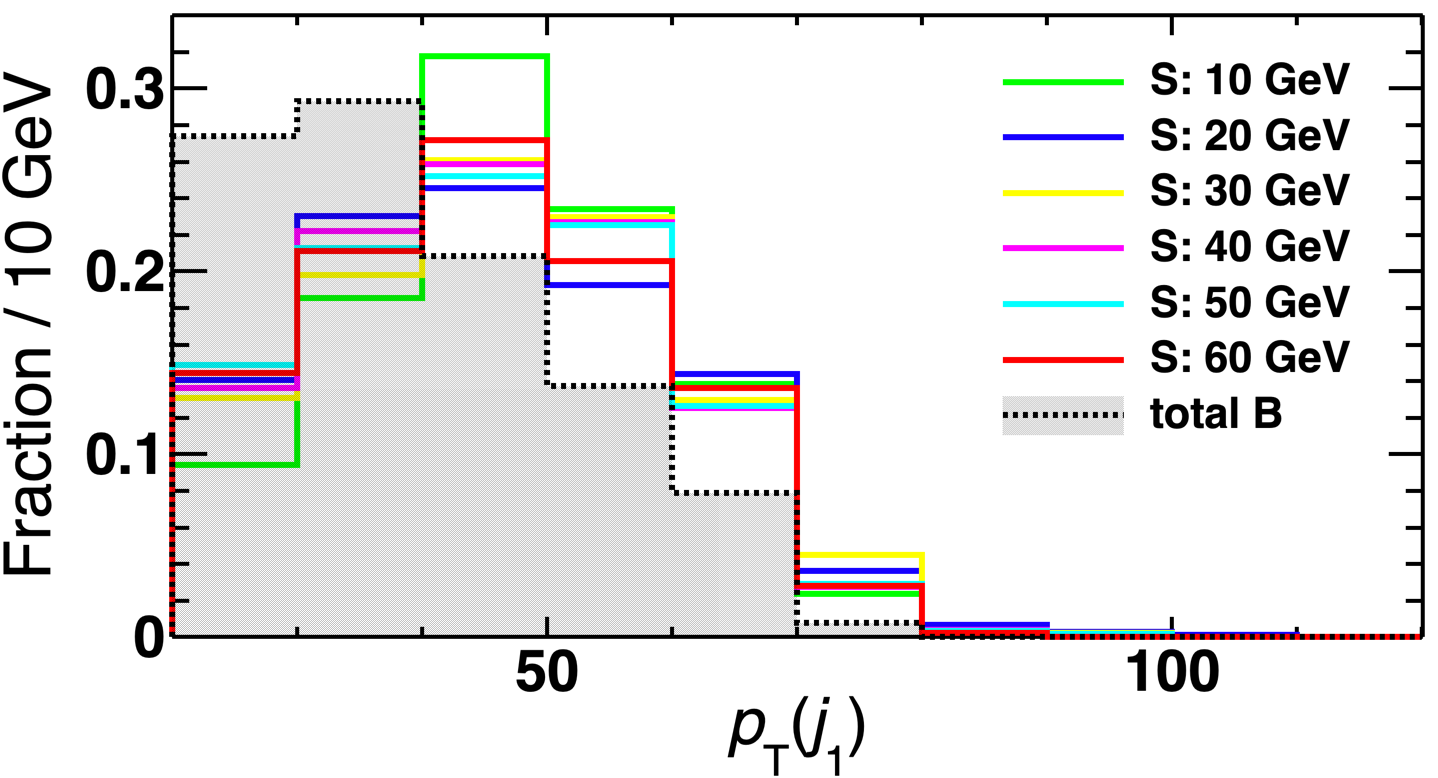}
\caption{
Normalized distribution of selected kinematic variables for the $4\ell$ channel. 
$p_T(\ell)$ panels are after selecting $N(l)=4$, $N(j)$ is after cuts(i-ii), $p_T(j)$ panels are after cuts(i-iv).
}
\label{fig:Obs4l}
\end{figure} 

The kinetic distributions of selected observables for the signal with different $N$ masses and the total background are shown in Fig.~\ref{fig:Obs4l}. $p_T(\ell_1)$, $p_T(\ell_2)$, $p_T(\ell_3)$, $p_T(\ell_4)$ correspond to the samples after requiring $N(\ell)=4$; $N(j)$ is after selecting cuts(i-ii); $p_T(j_1)$ is after requiring cuts(i-iv). 

\begin{table}[h]
\centering
\begin{tabular}{c|c|cccc}
\hline
\hline
\multicolumn{2}{c|}{} & initial & cuts(i) & cuts(ii)& cuts(iii-iv)   \\
\hline
\multirow{6}{*}{Sig.} 
& 10 GeV  & $10^3$ & 15.9 & 1.1 & 0.71 \\
& 20 GeV  & $10^3$ & 17.5 & 1.1 & 0.72 \\
& 30 GeV  & $10^3$ & 22.1 & 1.3 & 0.80 \\
& 40 GeV  & $10^3$ & 26.8 & 1.5 & 0.98 \\
& 50 GeV  & $10^3$ & 30.1 & 1.8 & 1.2 \\
& 60 GeV  & $10^3$ & 32.1 & 2.1 & 1.3 \\
\hline
\multirow{8}{*}{Bkg.}
& $4\tau$    & $1.69\times10^4$   & 58.4  & 6.8  & - \\
& $^\dagger 2\tau Z$ & $6.80\times10^5$   & $2.26\times10^3$  & 9.6  & -  \\
& $^\dagger 2\ell Z$  & $1.74\times10^6$   & $7.28\times10^4$  & - & -  \\
& $4\tau Z$ & 93.0   & 0.45  & $6.4\times10^{-3}$   & $2.8\times10^{-3}$  \\
& $2\tau 2W$ & $4.42\times10^3$  & 1.3  & 0.17  & - \\
& $^\dagger 2\ell 2\tau Z$ &  584  & 13.8 & $1.0\times10^{-2}$  & $3.2\times10^{-3}$  \\
& $^\dagger 4\ell Z$ & 862 &  116 & $7.8\times10^{-4}$  & - \\
& $^\dagger 2\ell 2W$ & $2.74\times10^4$ & 217 & -  & -  \\
\hline
\hline
\end{tabular}
\caption{
Similar to Table~\ref{tab:2l} but for the 4$\ell$ channel.
Background channels with $^\dagger$ require missing leptons or wrong signs.
}
\label{tab:4l}
\end{table}

The expected number of events at different cut stages for signal with different $N$ masses and background channels are shown in Table~\ref{tab:4l}. Due to two SSSF dileptons, two wrong-sign leptons must occur to fake such an event. Missing leptons are also less a problem as it would take a $2\tau 2e2\mu Z$ final state with one missed $e$ and one missed $\mu$ to fake the signal.

The surviving backgrounds are the $4\tau Z$ and $2\ell 2\tau Z$ channels. The lepton flavor and opposite-sign cuts play the central role in rejecting the background with same flavor, oppose-sign leptons. Stringent lepton counting removes the background from hadronic $\tau$ decays. The $2\tau 2\ell Z$ channel still contributes to background events, possibly due to wrong sign leptons.

\medskip
\section{Results}
\label{sect:summary}

Signal event samples are generated with the 
$e^-e^+\to Z h_1 \to Z (N N)$
process and then we let $Z$-boson and the heavy neutrinos decay inclusively. The signal event rate is,
\be 
N_s = L \cdot \sigma_{Zh_1} \cdot {\rm BR(}h_1\rightarrow NN)\cdot \eta_{\rm s}
\label{eqn:Ns}
\ee
where $L$ is the collider luminosity and $\eta_{\rm s}$ denotes the cut-based selection efficiency on the generated signal events. 
\begin{figure}[H]
\centering
\includegraphics[width=8.5cm,height=5.5cm]{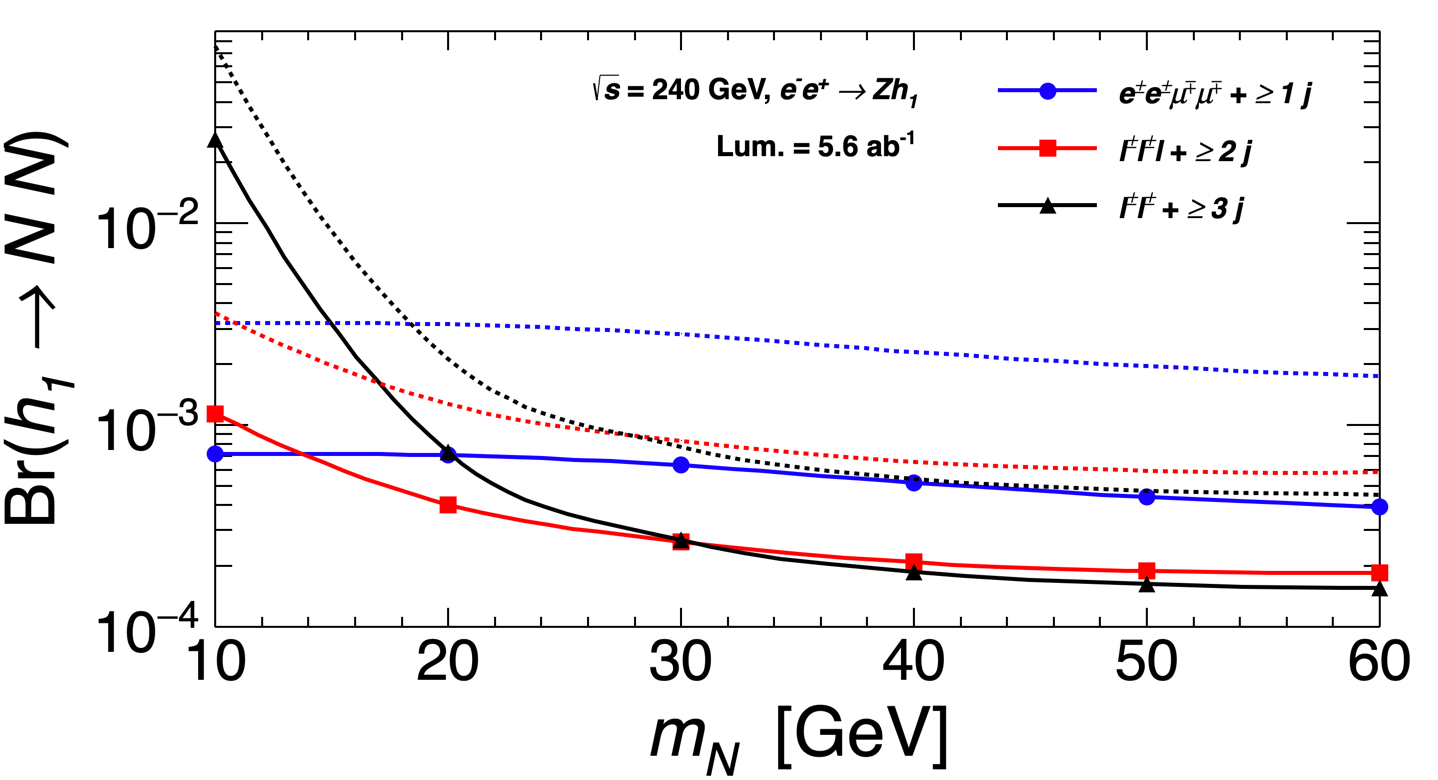}
\caption{
Sensitivity limits on the decay branching ratio of Higgs boson to $NN$ for 2-4$\ell$ channels assuming $m_N$ between 10 and 60 GeV.
$Zh_1$ production assumes 240 GeV center-of-mass energy and 5.6 ab$^{-1}$ integrated luminosity at future $e^-e^+$ colliders.
The solid (dotted) curves correspond to $2\sigma$($5\sigma$) significance.
}
\label{fig:limits}
\end{figure}
With a design luminosity $L=5.6$ ab$^{-1}$ and $\sigma_{Zh_1}=196$ fb at 240 GeV center-of-mass energy~\cite{CEPCStudyGroup:2018ghi}, a sample of $1.1\times 10^6$ $Zh_1$ events are expected. Since we let $N$ decay inclusively, $\eta_s$ already includes the $NN$ system's combined branching fraction into the selected final states, thus the formula above does not explicitly contain the $N$ decay branching fraction. The signal's statistic significance is
\be 
\sigma_{\rm stat} = \sqrt{2 [(N_s+N_b) {\rm ln}(1+\frac{N_s}{N_b}) - N_s ] }.
\ee
The background event rates $N_{\rm b}$ for $2\ell, 3\ell$ and $4\ell$ channels are listed in Table ~\ref{tab:2l},~\ref{tab:3l} and~\ref{tab:4l}, respectively. Requiring $2\sigma$ and $5\sigma$ significance levels, the sensitivity limits on BR($h_1\rightarrow NN$) are shown in Fig.~\ref{fig:limits}.

Given $L\cdot \sigma_{Zh_1} \sim 10^6$, the BR($h_1\rightarrow NN$)$\eta_s$ combination in Eq.~(\ref{eqn:Ns}) is statistically limited by $N_s/N_{Zh_1}\sim 10^{-6}N_s$. The selection efficiency $\eta_s$ is favorably evaluated via Monte Carlo, as $\eta_s$ is weighted between different decay chains that contribute to the same final state. In our case, both $Z\rightarrow \ell\ell, jj$ contribute to the signal channels. 

Not that $2\ell$ and $3\ell$ limits worsen towards lower $m_N$. This is caused by $N$ decaying into collimated leptons and jets when $N$ becomes more boosted at smaller $m_N$, resulting in fewer and softer reconstructed jets, hence hard hit by the jet selection cuts. Relaxing the jet cuts would help recovering more low $m_N$ signal events, yet at the cost of significantly increasing the SM background. The $4\ell$ channel has the highest background veto efficiency that can saturate luminosity cap ($N_b<1$) up to $10^3~\iab$. Despite the sub-unity background event rate, the $4\ell$ channel's sensitivity is less stringent than $2\ell$ and $3\ell$ channels for larger $m_N$ because of its much lower signal selection efficiency.

BR($h_1\rightarrow NN$) relates to BSM parameters as
\begin{eqnarray}
\left| \sin\alpha\cdot y_S \right|^2
&=& {\rm BR}(h_1\rightarrow NN) \nonumber \\ 
&\times & 16\pi \frac{\Gamma_{h_1}}{m_{h_1}} \left(1-\frac{4m_N^2}{m_{h_1}^2} \right)^{-3/2}.
\label{eqn:sinAlpha}
\end{eqnarray}
A BR($h_1\rightarrow NN$)$=10^{-4}$ sensitivity limit would correspond to $|\sin{\alpha}\cdot y_S|^2\le 7.3\times 10^{-6}$ at $m_N=60$ GeV, or $\le 1.6\times 10^{-7}$ towards low $N$ mass $2m_N \ll m_{h_1}$, where the decay phase space is least suppressed by mass gap as $\left(1-\frac{4m_N^2}{m_{h_1}^2}\right)\rightarrow 1$ in Eq.~(\ref{eqn:sinAlpha}). 

$y_S$ is a free model parameter given by $v_S$ and $m_N$. For $y_S\sim {\cal O}(1)$, this limit constrains $|\sin \alpha|$ to be lower than $10^{-2}$. This shows that the future Higgs factory has good sensitivity to tiny effective mixing angles between the Higgs boson and the BSM singlet scalar, which is comparable to the projected $|\sin\alpha|^2\sim 10^{-4}$ sensitivity at the LHC~\cite{Gao:2019tio}.

\begin{acknowledgments}
\noindent
We thank Qin Qin and Manqi Ruan for the useful discussions.
Y.G. is supported by the Institute of High Energy Physics, Chinese Academy of Sciences under the CEPC theory grant (2019-2020) and partially under grant no. E2545AU210. 
K.W. is supported by the National Natural Science Foundation of China under grant no.~11905162, 
the Excellent Young Talents Program of the Wuhan University of Technology under grant no.~40122102, the research program of the Wuhan University of Technology under grant no.~2020IB024, and the CEPC theory grant (2019--2020) of the Institute of High Energy Physics, Chinese Academy of Sciences.
\end{acknowledgments}

\bibliography{Refs}
\bibliographystyle{h-physrev5}

\end{document}